\documentclass[prd,aps,a4,twocolumn,superscriptaddress,preprintnumbers,nofootinbib]{revtex4-1}
\usepackage{pslatex}
\usepackage[pdftex]{graphicx}
\usepackage{psfrag}
\usepackage{epsfig}
\usepackage{color}
\usepackage{cancel}
\usepackage{slashed}
\usepackage{amssymb}
\usepackage{amsmath}
\usepackage{hyperref}
\usepackage{enumerate}
\usepackage{multirow}
\begin{document}
\title{Extraction of neutron density distributions from
  high-statistics coherent elastic neutrino-nucleus scattering data}%
\author{D. Aristizabal Sierra}%
\email{daristizabal@ulg.ac.be}%
\affiliation{Universidad T\'ecnica
  Federico Santa Mar\'{i}a - Departamento de F\'{i}sica\\
  Casilla 110-V, Avda. Espa\~na 1680, Valpara\'{i}so, Chile}%
\begin{abstract}
  Forthcoming fixed-target coherent elastic neutrino-nucleus
  scattering experiments aim at measurements with
  $\cal{O}(\text{tonne})$-scale detectors and substantially reduced
  systematic and statistical uncertainties. With such high quality
  data, the extraction of point-neutron distributions mean-square
  radii requires a better understanding of possible theoretical
  uncertainties. We quantify the impact of single-nucleon
  electromagnetic mean-square radii on the weak-charge form factor and
  compare results from weak-charge form factor parametrizations and
  weak-charge form factor decompositions in terms of elastic vector
  proton and neutron form factors, including nucleon form factors
  $Q$-dependent terms up to order $Q^2$. We assess as well the
  differences arising from results derived using weak-charge form
  factor decompositions in terms of elastic vector proton and neutron
  form factors and a model-independent approach based solely on the
  assumption of spherically symmetric nuclear ground state. We
  demonstrate the impact of the main effects by assuming pseudo-data
  from a one-tonne LAr detector and find that, among the effects and
  under the assumptions considered in this paper, weak-charge form
  factor parametrizations and weak-charge form factor decompositions
  in terms of elastic vector proton and neutron form factors enable
  the extraction of the $^{40}\text{Ar}$ point-neutron distribution
  mean-square radius with a $\sim 15\%$ accuracy. With a substantial
  reduction of the beam-related neutron and steady-state backgrounds a
  $\sim 1\%$ precision extraction seems feasible, using either of the
  two approaches.
\end{abstract}

\maketitle
\tableofcontents
\section{Introduction}
\label{sec:intro}
Since its first observation by the COHERENT collaboration in 2017
\cite{Akimov:2017ade}, CE$\nu$NS has become a powerful tool for the
determination of nuclear and SM properties as well as for constraining
new physics. Current datasets involve measurements in CsI and LAr
detectors with fiducial volumes in the order of $10\,$kg
\cite{Akimov:2017ade,COHERENT:2021xmm,Akimov:2020czh}. With such
active volumes, the datasets comprise about $\sim 100$ events,
with---to a certain degree---sizable systematics and statistical
uncertainties. Using these data a wide range of analyses have been
carried out. They include the determination of cesium, iodide and
argon point-neutron distributions root-mean-square (rms) radii
\cite{Cadeddu:2017etk,Papoulias:2019txv,Miranda:2020tif,Cadeddu:2021ijh,Coloma:2020nhf,DeRomeri:2022twg},
measurements of the weak mixing angle at renormalization scales of the
order of 50 MeV
\cite{Papoulias:2017qdn,Papoulias:2019txv,Miranda:2020tif,DeRomeri:2022twg},
constraints on new interactions in the neutrino sector
\cite{Liao:2017uzy,AristizabalSierra:2018eqm,AristizabalSierra:2019ufd,AristizabalSierra:2019ykk,Dutta:2019eml,Cadeddu:2020nbr,Banerjee:2021laz,Abdullah:2018ykz,Shoemaker:2017lzs,Ge:2017mcq,Denton:2018xmq,Denton:2020hop,Coloma:2017ncl,Coloma:2019mbs},
constraints on neutrino electromagnetic properties
\cite{Cadeddu:2018dux,Miranda:2019wdy,Khan:2022akj} and limits on
sterile neutrinos \cite{Papoulias:2017qdn}.

Plans for enhancing statistics by deploying an
$\mathcal{O}(\text{tonne})$ LAr detector in the future second target
station (STS) at the Oak Ridge National Laboratory have been discussed
in Refs. \cite{ref:STS,Asaadi:2022ojm} (see
Ref. \cite{COHERENT:2019kwz} as well)\footnote{The Coherent
  Captain-Mills Experiment at Los Alamos National Laboratory aims at
  measurements with an $\mathcal{O}(\text{tonne})$ LAr detector as
  well \cite{captain-mills} (see
  e.g. Refs. \cite{CCM:2021leg,CCM:2021yzc} for discussions of its
  physics reach).}. With improvements on statistical and systematic
uncertainties, such detector will deliver high quality data with which
improvements on the different analyses that have been done so far are
expected. This calls for a better understanding of theoretical
uncertainties. For instance---depending on the experimental
uncertainties---the inclusion of one-loop order electroweak
corrections \cite{Tomalak:2020zfh} might become necessary, regardless
of the physics case the data will be used for.

As far as we know, the extraction of point-neutron distributions
mean-square radii with existing or forecasted data has been done
considering only leading order (LO) effects. With this in this paper
we mean the following: The measured nuclear charge radius and the
point-proton distribution rms radius have been assumed to be equal,
and only LO nucleon form factor terms ($Q$ independent terms) have
been considered
\cite{Kosmas:2017tsq,AristizabalSierra:2019zmy,Miranda:2020tif,AristizabalSierra:2021uob}. Furthermore,
analyses using nuclear weak-charge form factor parametrizations,
nuclear weak-charge form factor decompositions in terms of elastic
vector proton and neutron form factors and power series expansions of
the nuclear weak-charge form factor have been used
\cite{Patton:2012jr,Cadeddu:2017etk,Coloma:2020nhf,AristizabalSierra:2019zmy,AristizabalSierra:2021uob}. With
all of them implying---in principle---different precision levels.

In this paper we quantify the uncertainties implied by considering
only LO effects---defined as we have done in the previous
paragraph---in the extraction of point-neutron distributions
mean-square radii. The analysis is split in two parts. In the first
part, we quantify uncertainties at the nuclear weak-charge level for
heavy and light nuclei (cesium and argon, taken as representative
examples of heavy and light nuclei). To do so we compare results from
calculations using only LO effects with calculations where: (a) The
point-proton distribution radius is corrected with single-nucleon
electromagnetic mean-square radii, (b) nucleon form factor
$Q$-dependent terms (up to order $Q^2$) are included \footnote{Other
  subleading effects involve as well relativistic Darwin-Foldy and
  spin-orbit corrections \cite{Friar:1997js,Horowitz:2012we}, which
  our analysis does not account for.}. We then determine the
uncertainties implied by the most commonly used form factor
parametrizations and by expansions in terms of even moments.

In the second part, we quantify the precision at which the
point-neutron distribution rms radius can be extracted, including the
main effects found in the first part. In this case we proceed by
assuming pseudo-data from a one-tonne LAr detector with assumptions on
the beam-related neutron (BRN) and steady-state (SS) backgrounds as
well as systematic uncertainties extrapolated from current COHERENT
measurements.

The remainder of this paper is organized as follows. In
Sec. \ref{sec:charge-and-weak-FFs} we provide a general discussion of
the nuclear charge and weak-charge form factors as well as of the
nuclear charge and weak-charge radii. In
Sec. \ref{sec:quantitative-differences} we determine nuclear
weak-charge form factor uncertainties, while in
Sec.~\ref{eq:cevns-xsec} we briefly discuss the CE$\nu$NS differential
cross section and differential event rate. In
Sec. \ref{sec:neutron-distribution-limits}, based on the results from
Sec. \ref{sec:quantitative-differences}, we extract the
$^{40}\text{Ar}$ neutron distribution mean-square radius. Our
conclusions are presented in Sec. \ref{sec:conclusions}
\section{Nuclear charge and weak-charge form factors}
\label{sec:charge-and-weak-FFs}
The single-nucleon electromagnetic and weak currents can be expressed
in terms of the Sachs form factors \cite{Ernst:1960zza}. For a nucleon
$N$ ($N=n,p$) they read
\begin{widetext}
\begin{align}
  \label{eq:EM-currents}
  J_\text{EM}^{\mu,N}&=\overline{u}^{s^\prime}(p^\prime)
                   \left[
                   G_E^N\gamma^\mu + \left(\frac{G_M^N - G_E^N}{1+\tau}\right)
                   \left(\tau \gamma^\mu + i\sigma^{\mu\nu}\frac{q_\nu}{2m}\right)
                   \right]
                   u^s(p)\ ,
  \\
  \label{eq:NC-currents}
  J_\text{NC}^{\mu,N}&=\overline{u}^{s^\prime}(p^\prime)\left[\widetilde{G}_E^N\gamma^\mu
  + \left(\frac{\widetilde{G}_M^N - \widetilde{G}_E^N}{1+\tau}\right)
  \left(\tau \gamma^\mu + i\sigma^{\mu\nu}\frac{q_\nu}{2m}\right)\right]u^s(p)\ .
\end{align}
\end{widetext}
where $m$ is a universal nucleon mass, $q=p^\prime-p$ the transferred
four momentum and $\tau=Q^2/4/m^2$ (with $Q^2=-q^2$ a timelike vector)
and $u^s(p)$ and $u^{s^\prime}(p^\prime)$ on-shell nucleon
spinors. $G_{E,M}^N=G_{E,M}^N(Q^2)$ and
$\widetilde{G}^N_{E,M}=\widetilde{G}^N_{E,M}(Q^2)$ are the
single-nucleon form factors for $N$. The weak neutral current (NC)
involves as well axial and pseudoscalar form factors, which are
sensitive to the nucleon spin distribution in the nuclear medium. For
elastic scattering their contribution can then be regarded as a
subleading effect (actually vanishing in nuclei with even number of
protons and neutrons), and so are not considered in
Eq. (\ref{eq:NC-currents}).

The nuclear charge and weak-charge form factors follow from: (i)
Trading the on-shell nucleon spinors to spinor wavefunctions that
account for nucleons in a potential, (ii) assuming the impulse
approximation (i.e. assuming that single-nucleon form factors are
valid as well in the nuclear medium), (iii) integration over nuclear
volume, (iv) taking into account contributions from protons and
neutrons (for details see e.g. \cite{Horowitz:2012we}). Explicitly one
finds
\begin{align}
  \label{eq:EM-FFs}
  ZF_\text{C}&=
  \sum_{N=p,n}
  \left\{
    G_E^NF_V^N
  + \left(\frac{G_M^N 
  - G_E^N}{1+\tau}\right)
  \left(\tau F_V^N + \frac{q}{2m}F_T^N\right)
  \right\}\ ,
  \\
  \label{eq:EW-FFs}
  Q_\text{W}F_\text{W}&=
  \sum_{N=p,n}
  \left\{\widetilde{G}_E^NF_V^N 
  + \left(\frac{\widetilde{G}_M^N 
  - \widetilde{G}_E^N}{1+\tau}\right)
  \left[\tau F_V^N + \frac{q}{2m}F_T^N\right]\right\}\ ,
\end{align}
with $Q_W$ determined by the couplings of the proton and neutron to
the $Z$ gauge boson, $Q_\text{W}=N\,g_V^n + Z\,g_V^p$
($g_V^p=1/2-2\sin^2\theta_W$ and $g_V^n=-1/2$) \footnote{One-loop
  corrected proton and neutron electroweak couplings are given by
  $g_V^p=0.0721$ and $g_V^n=-0.988$ \cite{Liu:2007yi}. These are the
  values we use in the calculations presented in
  Secs. \ref{sec:quantitative-differences} and
  \ref{sec:neutron-distribution-limits}.}. Note that the form factors
written as in Eqs. (\ref{eq:EM-FFs}) and (\ref{eq:EW-FFs}) satisfy the
normalization condition $F_\text{C}(Q^2=0)=F_\text{W}(Q^2=0)=1$. These
expressions contain information on nucleon and nuclear structure, the
latter encoded in the vector and tensor nuclear form
factors. Spin-orbit effects governed by $F_T$ are subleading compared
with nuclear spin-independent effects controlled by $F_V$
\cite{Horowitz:2012we}. Thus keeping only LO effects and taking into
account that $\tau\ll 1$ for the typical transferred momentum in
neutrino stopped-pion sources ($Q^2\lesssim m_\mu^2/2$), the charge
and weak-charge form factors reduce to a rather simple form
\begin{align}
  \label{eq:EM-FFs-reduced-form}
  F_\text{C}(q^2)&=\frac{1}{Z}
  \left[G_E^p(q^2) F^p_V(q^2) + G_E^n(q^2) F^n_V(q^2)\right]\ ,
  \\
  \label{eq:EW-FFs-reduced-form}
  F_\text{W}(q^2)&=\frac{1}{Q_\text{W}}
 \left[\widetilde{G}_E^p(q^2) F^p_V(q^2) 
  + \widetilde{G}_E^n(q^2) F^n_V(q^2)\right]\ .
\end{align}
Numerically one finds that spin-orbit effects are of the order of
$0.1\%$ \cite{Horowitz:2012we}, while terms proportional to $\tau$
contribute at the order of $0.01\%$ in stopped-pion neutrino
experiments where $Q^2\simeq (10\,\text{MeV})^2$.
Eqs. (\ref{eq:EM-FFs-reduced-form}) and (\ref{eq:EW-FFs-reduced-form})
are thus precise enough at the percent level.
\subsection{Electromagnetic and weak charge radii}
\label{sec:charge-and-weak-radii}
LO expressions for the electromagnetic and weak-charge radii of the
nucleus follow from Eqs. (\ref{eq:EM-FFs-reduced-form}) and
(\ref{eq:EW-FFs-reduced-form}), as we now discuss. Assuming spherical
symmetry, the nucleon and spin-independent nucleon structure functions
can be expanded in terms of their moments. For nucleons one has
\begin{alignat}{2}
  \label{eq:nucleon-EMFF-expansion-moments}
  G_E^{p, n}&=\sum_{i=0, 1}^\infty(-1)^i\frac{Q^{2i}}{(2i+1)!}r_X^{2i}\ ,
  \\
  \label{eq:nucleon-EWFF-expansion-moments}
  \widetilde{G}_E^X&=g_V^X\sum_{i=0}^\infty(-1)^i
  \frac{Q^{2i}}{(2i+1)!}\tilde{r}_X^{2i}\ ,
\end{alignat}
where the order $Q^2$ terms in the expansion in
Eq. (\ref{eq:nucleon-EMFF-expansion-moments}) involve the
single-nucleon electromagnetic mean-square radii, $r_X^2$ ($X=p,n$),
while those in Eq. (\ref{eq:nucleon-EWFF-expansion-moments}) the
single-nucleon weak-charge mean-square radii, $\tilde{r}_X^2$. In
Eq. (\ref{eq:nucleon-EMFF-expansion-moments}) the sum starts at $i=0$
($i=1$) for protons (neutrons). For the spin-independent nuclear form
factors one instead has (see e.g. \cite{Patton:2012jr})
\begin{align}
  \label{eq:FF-nuclear-expansion-p}
  F_V^p&=Z\sum_{i=0}^\infty(-1)^i\frac{Q^{2i}}{(2i+1)!}R_p^{2i}\ ,
  \\
  \label{eq:FF-nuclear-expansion-n}
  F_V^n&=N\sum_{i=0}^\infty(-1)^i\frac{Q^{2i}}{(2i+1)!}R_n^{2i}\ .
\end{align}
In this case---at order $Q^2$---terms involve the point-proton and
point-neutron distributions mean-square radii, $R_p^2$ and $R_n^2$
(second moment of the distributions)\footnote{Here following standard
  conventions we use lower-case for nucleons and upper-case for
  nuclei.}. The nuclear charge and weak-charge radii, $R_\text{C}^2$
and $R_\text{W}^2$, follow from
\begin{equation}
  \label{eq:derivative_C_EW_charge_radii}
  R_\text{C}^2=-\left .6\frac{dF_\text{C}}{dQ^2}\right|_{Q^2=0}\ ,\quad
  R_\text{W}^2=-6\left .\frac{dF_\text{W}}{dQ^2}\right|_{Q^2=0}\ ,
\end{equation}
after expansions in Eqs. (\ref{eq:nucleon-EMFF-expansion-moments}),
(\ref{eq:nucleon-EWFF-expansion-moments}),
(\ref{eq:FF-nuclear-expansion-p}) and
(\ref{eq:FF-nuclear-expansion-n}) are inserted in
Eqs. (\ref{eq:EM-FFs-reduced-form}) and
(\ref{eq:EW-FFs-reduced-form}).  Their explicit expressions are given
by
\begin{align}
  \label{eq:EM-EW-charge-radii}
  R_\text{C}^2&=R_p^2 + r_p^2 + \frac{N}{Z}r_n^2\ ,
  \\
  R_\text{W}^2&=\frac{Zg_V^p}{Q_\text{W}}\left(R_p^2 + \tilde{r}_p^2\right)
                 +
                 \frac{Ng_V^n}{Q_\text{W}}\left(R_n^2 + \tilde{r}_n^2\right)\ .
\end{align}
The single-nucleon weak-charge mean-square radii in $R_\text{W}$ can
be reexpressed in terms of the single-nucleon electromagnetic
mean-square radii as follows \cite{Horowitz:2012we}
\begin{equation}
  \label{eq:weak-in-terms-of-EM}
  \tilde{r}_p^2=r_p^2 + \frac{g_V^n}{g_V^p}r_n^2 + \xi_V^{(0)}r_s^2\ ,
  \qquad
  \tilde{r}_n^2=r_p^2 + \frac{g_V^p}{g_V^n}r_n^2 + \xi_V^{(0)}r_s^2\ ,
\end{equation}
where $\xi_V^{(0)}=g_V^u+g_V^d+g_V^s$ (with $g_V^q$ the quark $q$
weak-vector charge) and $r_s^2$ the mean-square strange
radius. Numerically, $\xi_V^{(0)}=-0.988$ and
$r_s^2=-0.00430\,\text{fm}^2$ with the latter obtained from lattice
QCD calculations \cite{Sufian:2016pex}. Thus, the strange quark
contribution provides per mille corrections to the proton and neutron
LO expressions in Eq. (\ref{eq:weak-in-terms-of-EM}). It can therefore
be neglected in the following analysis.

With the aid of these equations the weak-charge mean-square radius can
finally be rewritten as
\begin{align}
  \label{eq:EW-mean-sqaure-rrius-final-exp}
  R_\text{W}^2&=\frac{Z}{Q_\text{W}}
 \left(g_V^p R_p^2 + g_V^p r_p^2 + g_V^n r_n^2\right)
\nonumber\\
  &+\frac{N}{Q_\text{W}}
 \left(g_V^n R_n^2 + g_V^n r_p^2 + g_V^p r_n^2\right)\ ,
\end{align}
which in turn can be recast in terms of the electromagnetic charge
radius and the neutron skin, $R_n^2-R_p^2$, \cite{Coloma:2020nhf}
\begin{equation}
  \label{eq:EW-charge-radius-neutron-skin}
  R_\text{W}^2=R_\text{C}^2 + \frac{Ng_V^n}{Q_\text{W}}
  \left[(R_n^2-R_p^2) + \frac{Z^2-N^2}{N\,Z} r_n^2\right]\ .
\end{equation}

Due to the small transferred momentum, stopped-pion CE$\nu$NS
experiments are not particularly sensitive to nucleon structure. Thus
the nucleon form factors in Eqs. (\ref{eq:EM-FFs-reduced-form}) and
(\ref{eq:EW-FFs-reduced-form}) can be truncated at order $Q^2$,
resulting in the following expression for the weak-charge form factor
\cite{Coloma:2020nhf}
\begin{align}
  \label{eq:EW-FF-approx}
  F_\text{W}&\simeq \frac{1}{Q_\text{W}}
               \left[
               Z\left(
               g_V^p 
               - \frac{g_V^p}{6}r_p^2 Q^2 
               - \frac{g_V^n}{6}r_n^2 Q^2
               \right) F_V^p(Q^2)\right .
               \nonumber\\
               &\left .
               +
               N\left(
               g_V^n 
               - \frac{g_V^n}{6}r_p^2 Q^2 
               - \frac{g_V^p}{6}r_n^2 Q^2
               \right) F_V^n(Q^2)
               \right]\ ,
\end{align}
where the nuclear form factors have been normalized, $F_V^p(Q^2=0)=1$
and $F_V^n(Q^2=0)=1$.
\section{Theoretical and phenomenological uncertainties}
\label{sec:quantitative-differences}
In this Section we determine the numerical variations to which the
weak-charge form factor can be subject to. For that aim we consider
the Helm parametrization \cite{Helm:1956zz} for the elastic vector
proton and neutron form factors (see Sec. \ref{sec:LO-exp}).  We first
quantify $F_\text{W}$ by considering LO expressions for the
point-proton distribution mean-square radius and the nucleon form
factors, which we compare with $F_\text{W}$ obtained by considering
the full expression in Eq. (\ref{eq:EW-FF-approx}), including the
single-nucleon electromagnetic mean-square radii in the determination
of $R_p$. We compare as well those results with those obtained
assuming the Helm parametrization for the weak-charge form factor. We
then proceed to calculate $F_\text{W}$ by assuming different
parametrizations for the elastic vector proton and neutron nuclear
form factors. Our analysis relies on the Fourier transform of the
symmetrized Fermi function \cite{Friar:1997js} and the Klein-Nystrand
\cite{Klein:1999qj} parametrizations, in addition to the Helm
approach. In doing so, we quantify the dependence of $F_\text{W}$ on
parametrization choice. Finally, we compare the parametrization
approach with the model-independent treatment based on series
expansions of the elastic vector proton and neutron form factors.
\begin{figure*}
  \centering
  \includegraphics[scale=0.385]{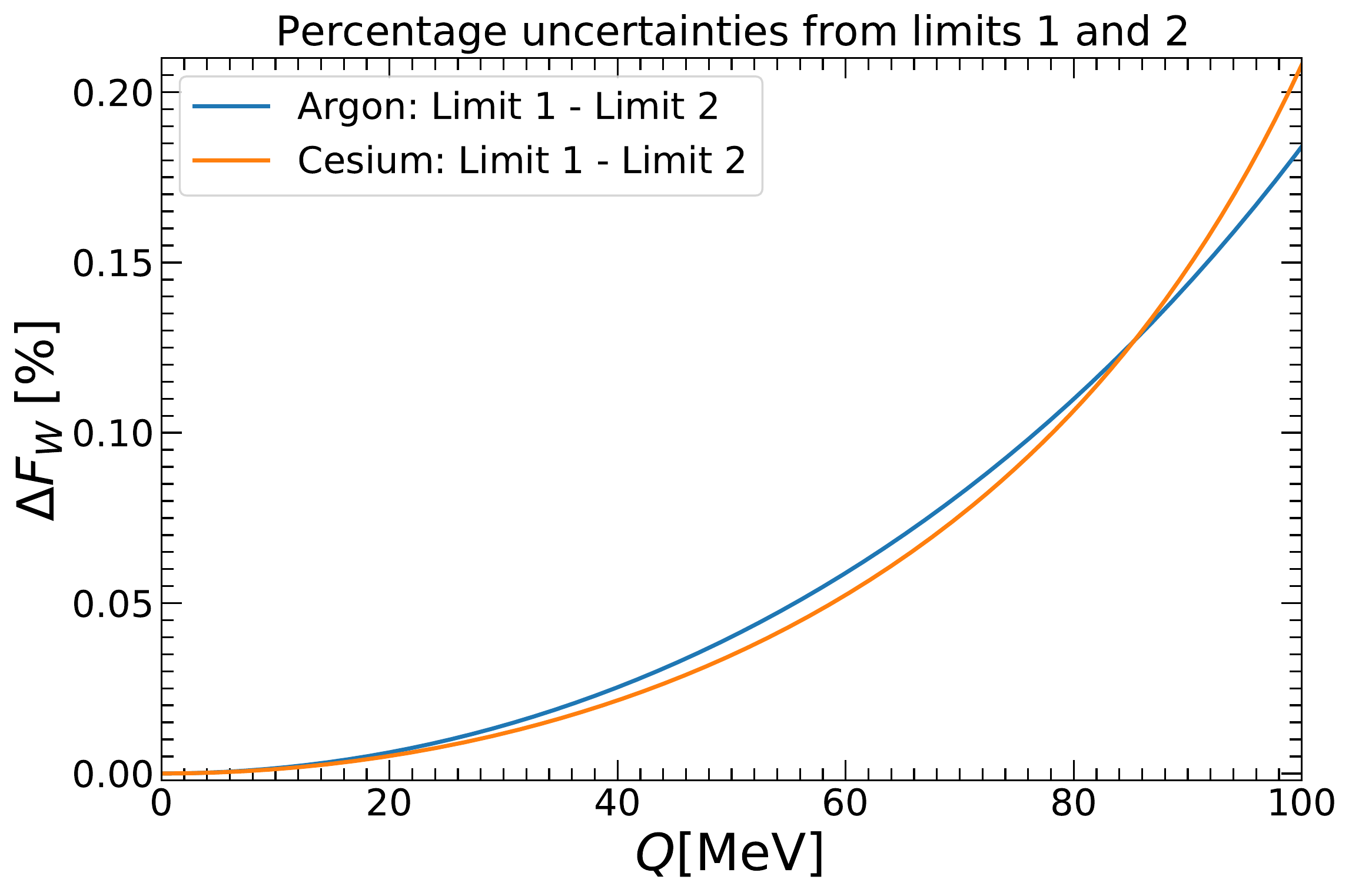}
  \includegraphics[scale=0.385]{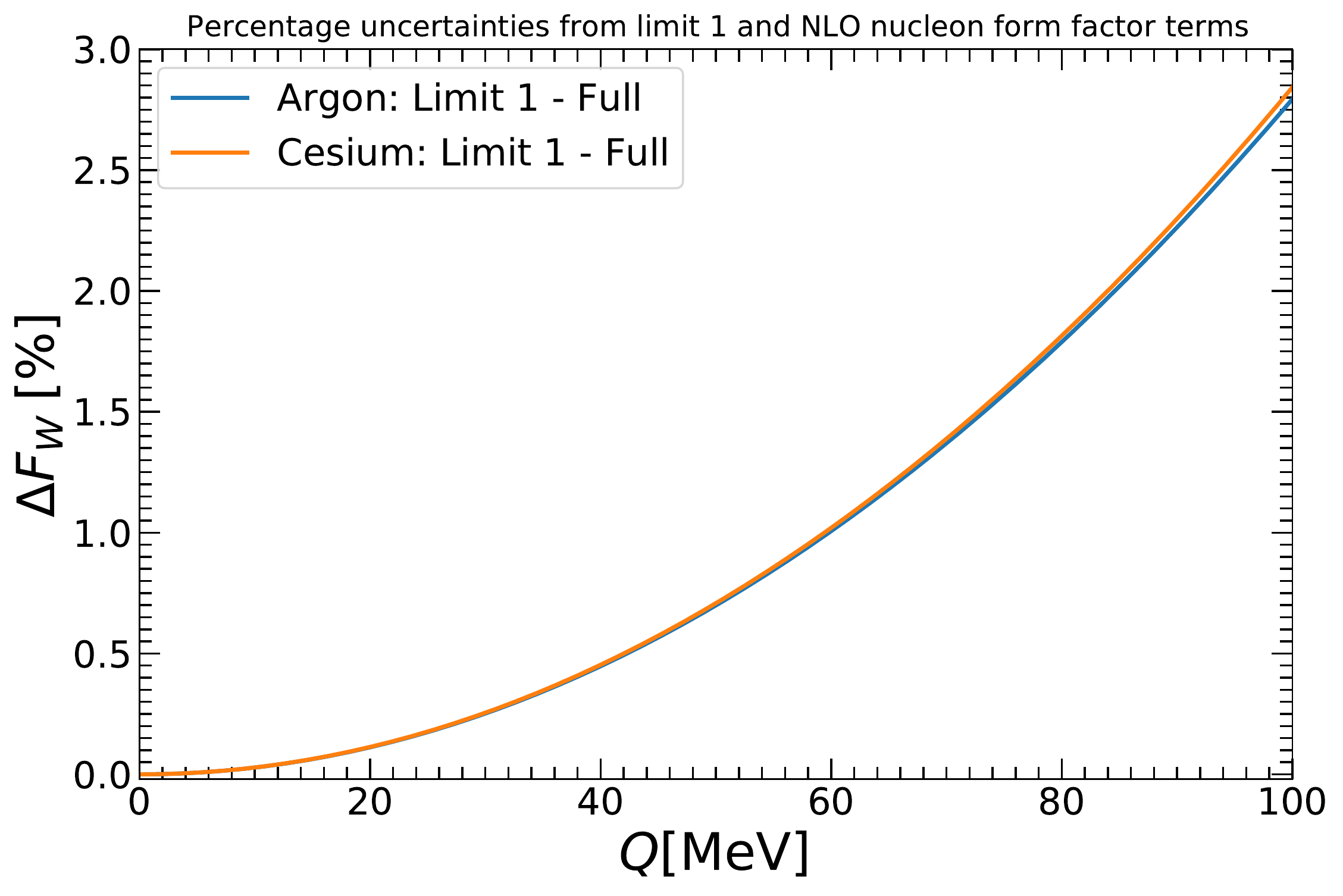}
  \caption{Nuclear weak-charge form factor percentage difference
    calculated according to Eq. (\ref{eq:percentage-diff}) and for:
    Nucleon form factors LO terms (momentum-independent terms only)
    and $R_p^2=R_\text{C}^2$ (\textbf{Limit 1}, item
    \ref{case1:limit-1}), nucleon form factor LO terms and
    $R_p^2=R_\text{C}^2-r_p^2-(N/Z) r_n^2$ (\textbf{Limit 2}, item
    \ref{case2:limit-2}), full nucleon form factors up to order $Q^2$
    (\textbf{Full}, item \ref{case3:limit-3}) and
    $R_p^2=R_\text{C}^2-r_p^2-(N/Z) r_n^2$. \textbf{Left graph}:
    Percentage difference obtained by comparing limits 1 and 2,
    \textbf{Right graph}: percentage difference obtained by comparing
    limit 1 with full nucleon form factor expressions and
    $R_p^2=R_\text{C}^2-r_p^2-(N/Z) r_n^2$. For the point-neutron
    distribution rms radii we have taken $R_n=R_p+0.1\,$fm and
    $R_n=R_p+0.2\,$fm for argon and cesium, respectively. These values
    taken just for the sake of illustrating the deviations implied by
    the different limits compared with the case in which full
    expressions are considered. For the $Q$ range displayed in the
    graphs the neutrino flux is sizable (see text for more details).}
  \label{fig:form-factor-PD-Ar-Cs}
\end{figure*}
\subsection{Uncertainties due to leading-order point-proton
  distribution mean-square radius and leading-order nucleon form
  factors}
\label{sec:LO-exp}
We start by considering three different cases, two limits and the full
case:
\begin{enumerate}[1)]
\item\label{case1:limit-1}\textbf{Limit 1 (Case 1)}: LO nucleon form
  factors ($Q$-independent terms) and point-proton distribution
  mean-square radius equal to the nuclear charge mean-square radius,
  $R_p^2=R_\text{C}^2$.
\item\label{case2:limit-2}\textbf{Limit 2 (Case 2)}: LO nucleon form
  factors and full point-proton distribution mean-square radius given
  by Eq. (\ref{eq:EM-EW-charge-radii}).
\item\label{case3:limit-3}\textbf{Full case (Case 3)}: Full nucleon form factors,
  up to $Q^2$ as given in Eq. (\ref{eq:EW-FF-approx}) and full
  point-proton distribution mean-square radius given by
  Eq. (\ref{eq:EM-EW-charge-radii}).
\end{enumerate}
These limits are motivated as follows. In most analyses in which the
point-neutron distribution rms radius is extracted from CE$\nu$NS data
(see e.g. Ref. \cite{AristizabalSierra:2021uob}) the point-proton
distribution mean-square radius is fixed to be equal to the nuclear
charge radius. As can be seen from Eq. (\ref{eq:EM-EW-charge-radii}),
doing so overestimates $R_p$. A larger value for $R_p$ means a smaller
value for $F_p$ at a given $Q$. Under such simplification, an
uncertainty in the extraction of $R_n^2$ from CE$\nu$NS data is
implied. Order $Q^2$ corrections in the nucleon form factors are also
typically ignored and, though suppressed, are potentially a source of
sizable uncertainties. Ignoring order $Q^2$ terms enhance $F_\text{W}$
[see Eq. (\ref{eq:EW-FF-approx})], and so the CE$\nu$NS event rate,
resulting in a potential underestimation of $R_n^2$.

To proceed, we adopt the Helm form factor parametrization for the
elastic vector proton and neutron form factors, $F_V^p$ and $F_V^n$,
namely \cite{Helm:1956zz}
\begin{equation}
  \label{eq:FF-Helm}
  F_H(Q^2)=3\frac{j_1(QR_0)}{QR_0}e^{-(Qs)^2/2}\ .
\end{equation}
Here $R_0$ refers to the diffraction radius, related with the
point-nucleon distribution mean-square radius through the skin
thickness $s=0.9\,$fm \cite{Lewin:1995rx} according to
\begin{equation}
  \label{eq:diffraction-radius-mean-square-radius}
  R_0=\sqrt{\frac{5}{3}\left( R_X^2 - 3s^2\right)}
  \qquad (X=p, n)\ .
\end{equation}

To avoid mixing nuclear physics effects with detector effects we use
argon and cesium as target materials. For argon this means one can
assume the detector to be $100\%$ composed of $^{40}\text{Ar}$ (up to
per mille corrections), while for cesium of
$^{133}\text{Cs}$. Parameters used in our calculation are displayed in
Tab. \ref{tab:argon-germanium-parameters}, along with the values for
the single-nucleon electromagnetic mean-square radii central values
\cite{Zyla:2020zbs}.
\begin{table}[h]
  \setlength{\tabcolsep}{12pt}
  \renewcommand{\arraystretch}{1.1}
  \centering
  \begin{tabular}{|c||c|c|}\hline
    Isotope & $^{40}\text{Ar}$ & $^{133}\text{Cs}$\\\hline
    Abundance ($X_i$) & 99.6\% & 100.0\% \\\hline
    $R_\text{C}$ [fm] & 3.4274 & 4.0414 \\\hline\hline
    Nucleon & Proton & Neutron\\\hline
    $r_X^2\,[\text{fm}^2]$  & 0.7071 & -0.1155 \\\hline
  \end{tabular}
  \caption{$^{40}\text{Ar}$ and $^{133}\text{Cs}$ relative abundances along 
    with their nuclear charge radii used in our calculation. 
    Nuclear charge radii taken from Ref. \cite{Angeli:2013epw}. Single-nucleon
    electromagnetic mean-square radii central values taken from Ref.
    \cite{Zyla:2020zbs}.}
  \label{tab:argon-germanium-parameters}
\end{table}

To quantify deviations implied by the limits in items
\ref{case1:limit-1} and \ref{case2:limit-2} with the full result in
item \ref{case3:limit-3} the following percentage difference factor is
employed
\begin{equation}
  \label{eq:percentage-diff}
  \Delta F_\text{W}\;[\%] = \frac{F_\text{W}|_{\text{C}_i} 
    - F_\text{W}|_{\text{C}_j}}{F_\text{W}|_{\text{C}_i}}\times 100\%\ ,
\end{equation}
where $F_\text{W}|_{\text{C}_i}>F_\text{W}|_{\text{C}_j}$ and
C$_{i,j}$ refer to the cases used for the calculation of the
weak-charge form factor. The results are shown in
Fig.~\ref{fig:form-factor-PD-Ar-Cs}. We have fixed the point-neutron
distribution rms radius according to $R_n=R_p+0.1\,$fm and
$R_n=R_p+0.2\,$fm for argon and cesium, respectively. Note that these
choices are not intended to represent the actual values. They are
rather used as proxies whose motivation is that of covering values
predicted by different theoretical nuclear methods
\cite{AristizabalSierra:2019zmy}. Results following from this choice
are rather stable provided $R_n$ does not largely differs from $R_p$,
somehow expected for the nuclei we are interested in.

From Fig. \ref{fig:form-factor-PD-Ar-Cs} (left graph) one can see that
by assuming that the point-proton rms radius distribution amounts to
the measured nuclear charge rms radius has a small effect. All over
the relevant transferred momentum range ($Q^2\lesssim m_\mu^2$)
deviations are at (or below) the per mille level, regardless of
nuclide. Including the single-nucleon electromagnetic mean-square
radii in the determination of $R_p^2$ becomes relevant only if
experimental uncertainties reach that level of precision, otherwise
the $R_p^2=R_\text{C}^2$ approximation is fairly accurate. Differences
due to nucleon form factors $Q$-dependent terms are instead slightly
more relevant. Results in Fig.~\ref{fig:form-factor-PD-Ar-Cs} (right
graph) show that they can reach values up to $3\%$, basically
regardless of nuclide. Nucleon form factors effects should then
accounted for in high-statistics CE$\nu$NS experiments with low
systematic and statistical uncertainties.

\begin{figure*}
  \centering
  \includegraphics[scale=0.385]{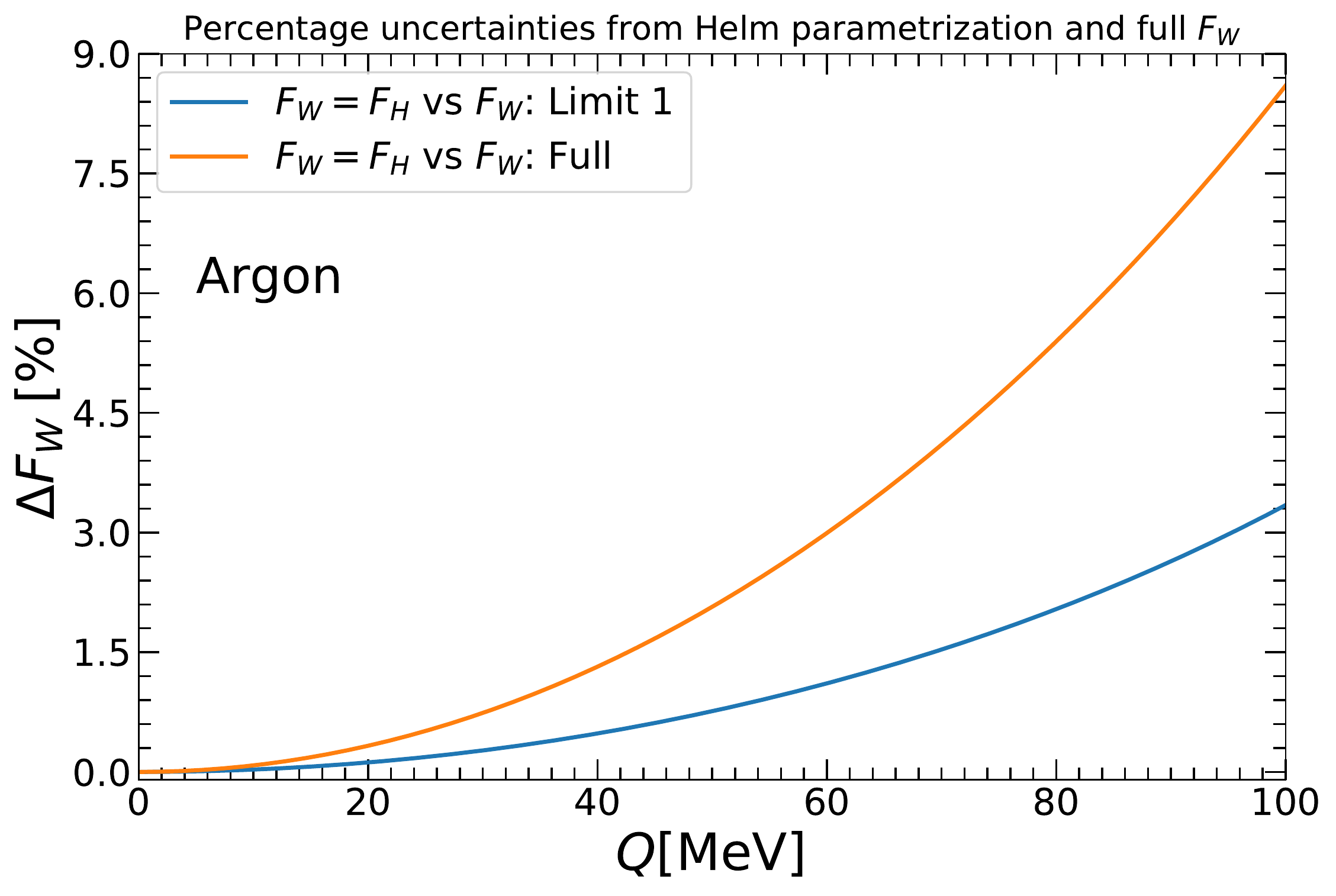}
  \includegraphics[scale=0.385]{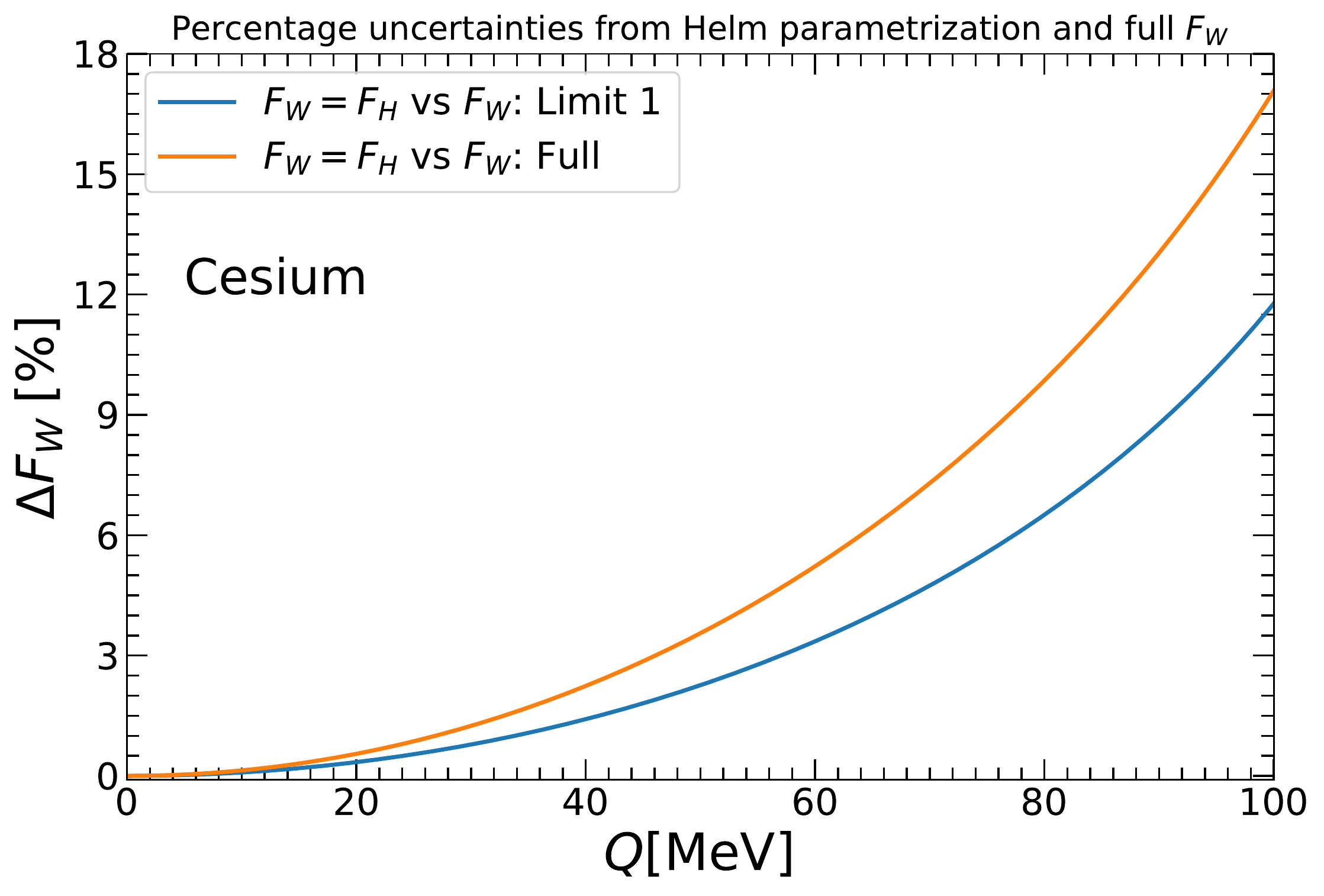}
  \caption{Nuclear weak-charge form factor percentage uncertainty for
    argon (left graph) and cesium (right graph) calculated using the
    Helm parametrization with the diffraction radius $R_0$ fixed
    through the weak-charge mean-square radius $R_\text{W}^2$ and by
    using Eq. (\ref{eq:EW-FF-approx}). Percentage uncertainties are
    calculated in limit 1 (item \ref{case1:limit-1}) and in the full
    case (item \ref{case3:limit-3}). The point-neutron distribution
    rms radii have been fixed according to $R_n=R_p + 0.1\,$fm and
    $R_n=R_p + 0.2\,$fm. For the $Q$ range displayed in the graphs the
    neutrino flux is sizable (see text for more details).}
  \label{fig:FW-two-equivalent-procedures}
\end{figure*}
Rather than calculating the weak-charge form factor using
Eq. (\ref{eq:EW-FF-approx}), one could instead use a form factor
parametrization for $F_\text{W}$---e.g. the Helm parametrization---and
fix $R_X^2=R_\text{W}^2$ in
Eq. (\ref{eq:diffraction-radius-mean-square-radius}), with
$R_\text{W}^2$ as given in
Eq. (\ref{eq:EW-charge-radius-neutron-skin}) (as done e.g. in
Ref. \cite{Coloma:2020nhf}). Since $R_\text{W}^2$ involves the
point-neutron distribution mean-square radius, from such procedure one
could as well extract from CE$\nu$NS data a range for its value at a
certain confidence level. To show what are the expected percentage
differences following this procedure with that dictated by
Eq. (\ref{eq:EW-FF-approx}), we have calculated $F_\text{W}$ as we
have described above and compared with $F_\text{W}$ calculated using
Eq. (\ref{eq:EW-FF-approx}) in limit 1 (item \ref{case1:limit-1}) and
in the full case (item \ref{case3:limit-3}). Deviations are quantified
with the aid of Eq. (\ref{eq:percentage-diff}) where in this case
C$_i$ refers to $F_\text{W}$ calculated through the Helm
parametrization and C$_j$ to $F_\text{W}$ calculated using
Eq. (\ref{eq:EW-FF-approx}) in the aforementioned cases.

The result is displayed in
Fig. \ref{fig:FW-two-equivalent-procedures}, left (right) graph for
argon (cesium). One can see that differences between both approaches
are more pronounced for heavier nuclei. And grow depending on whether
one considers the most simplified assumptions (limit 1) or the full
weak-charge form factor including nucleon form factors corrections and
single-nucleon electromagnetic mean-square radii. For argon, the
percentage uncertainty can raise up to $\sim 9\%$ while for
cesium---instead---up to $\sim 18\%$. Some caution, however, is
required. Differences grow with transferred momentum and so at their
peak (in the relevant range) the stopped-pion neutrino flux is
expected to be fading away. Because of kinemtic reasons, the
contribution of $\nu_e$ to the delayed component fades away at
$E_\nu=m_\mu/2$. The contribution of $\overline{\nu}_\mu$ has a sharp
fall there as well, but contributes sizably at energies close to the
kinematic cut. Thus, evaluating $E_r^\text{max}$ at $E_\nu=m_\mu/2$
and calculating with that value $Q_\text{max}$ for argon and cesium
one finds $Q_\text{max}^\text{Cs}\simeq 100\,$MeV and
$Q_\text{max}^\text{Ar}\simeq 190\,$MeV. It is for this reason that we
have plotted only up to $Q=100\,$MeV. To assure that deviations are
shown in the region where they play a role at the event rate level.

\begin{figure*}
  \centering
  \includegraphics[scale=0.385]{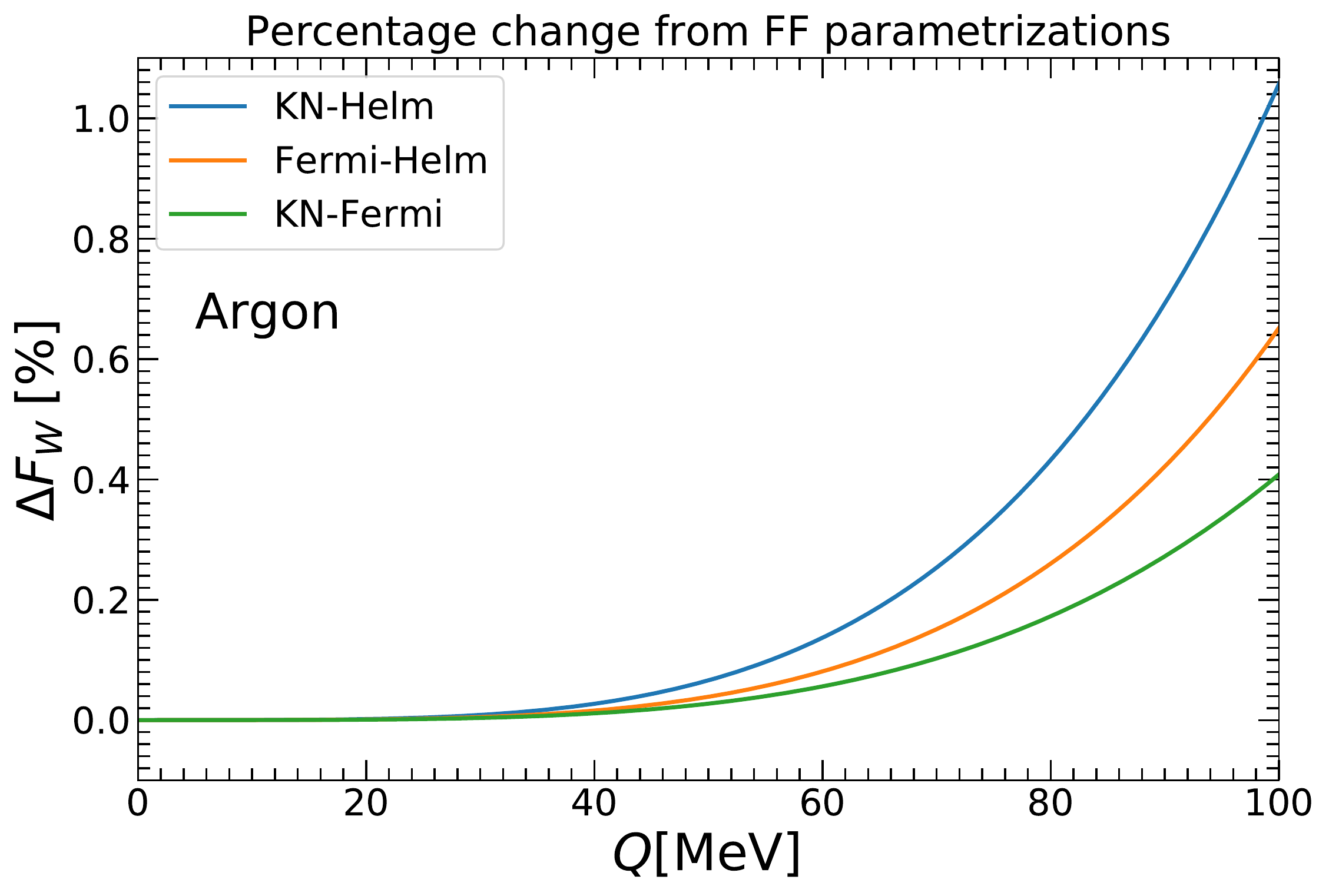}
  \includegraphics[scale=0.385]{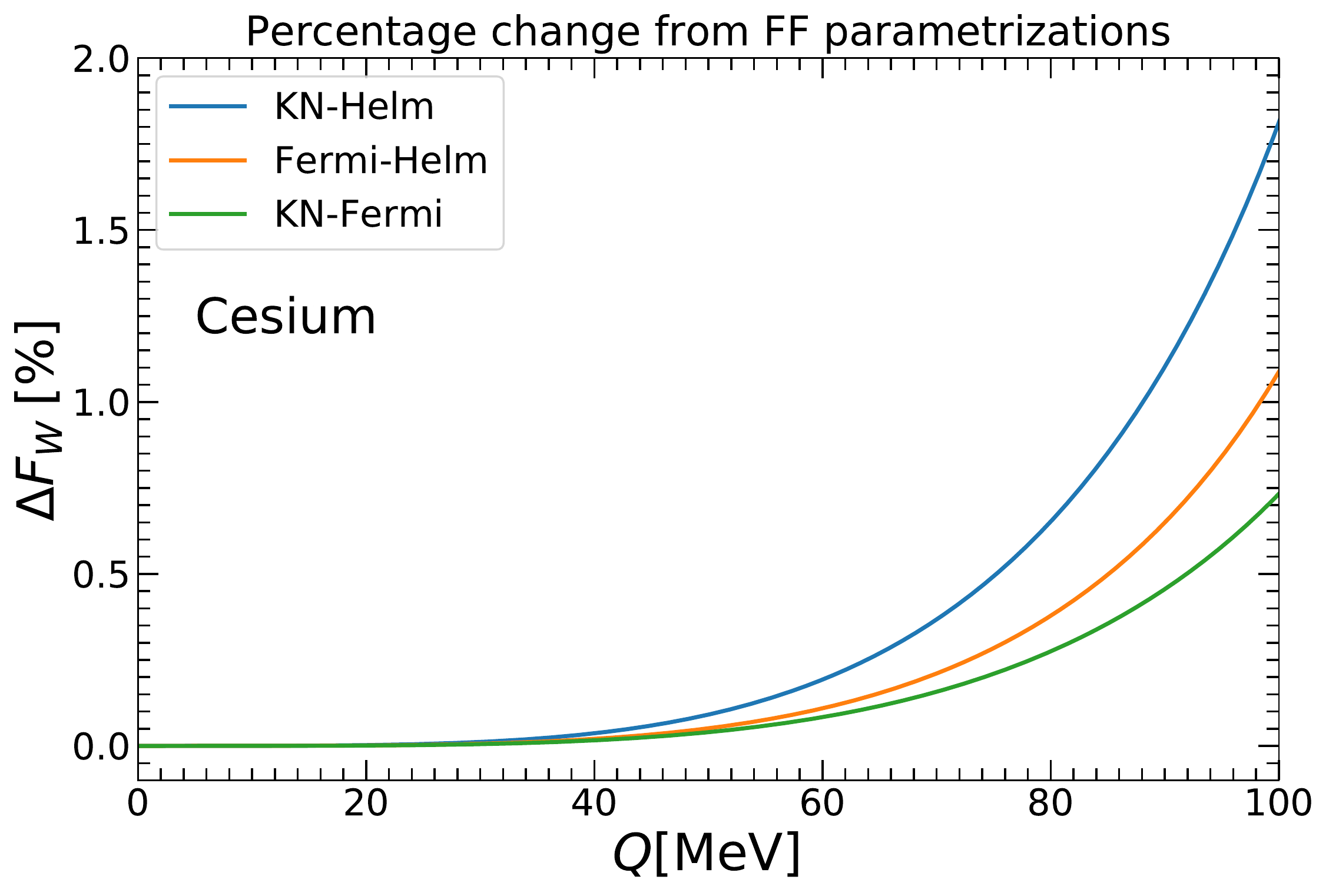}
  \caption{\textbf{Left graph}: Weak-charge form factor percentage
    difference for argon obtained by using for the elastic vector
    proton and neutron form factors the Helm, Klein-Nystrand and the
    Fourier transform of the symmetrized Fermi distribution
    parametrizations. The results include nucleon form factors
    $Q$-dependent terms (up to order $Q^2$) and single-nucleon
    electromagnetic mean-square radii. The argon point-neutron
    distribution rms radius has been fixed according to
    $R_n=R_p + 0.1\,$fm. \textbf{Right graph}: Same as left graph but
    for cesium, with the point-neutron distribution rms radius fixed
    according to $R_n=R_p + 0.2\,$fm. For the $Q$ range displayed in
    the graphs the neutrino flux is sizable (see text for more
    details).}
  \label{fig:form-factor-parametrization-dependences}
\end{figure*}
\subsection{Uncertainties due to elastic vector proton and neutron
  form factor parametrizations}
\label{sec:form_factor_param}
Event rate spectra derived from form factor parametrizations are
expected to depend on the parametrization used, with the dependence
increasing with increased $Q$ \cite{AristizabalSierra:2019zmy}. To
determine the size of these dependences we calculate $F_V^p$ and
$F_V^n$ using as well the Klein-Nystrand form factor
\cite{Klein:1999qj} and the Fourier transform of the symmetrized
Fermi distribution \cite{Friar:1997js}.

The Klein-Nystrand form factor follows from folding a Yukawa potential
(range $a_k$) over a hard sphere distribution of radius $R_A$, namely
\cite{Klein:1999qj}
\begin{equation}
    \label{eq:KN-FF}
    F_\text{KN}=3\frac{j_1(QR_A)}{QR_A}\frac{1}{1+Q^2a_k^2}\ .
\end{equation}
The range of the potential $a_k$ is 0.7 fm \cite{Klein:1999qj} and the
hard sphere radius is determined through the point-proton and
point-neutron distributions mean-square radii according to
\begin{equation}
  \label{eq:radius-potential-range-KN-FF}
  R_A=\sqrt{\frac{5}{3}\left(R^2_X - 6a_k^2\right)}
  \qquad (X=p, n)\ .
\end{equation}
The Fourier transform of the symmetrized Fermi distribution follows
instead from $f_\text{SF}=f_\text{F}(r)+f_\text{F}(-r)-1$, where
$f_\text{F}(r)$ is the conventional Fermi function
\begin{equation}
  \label{eq:fermi-function}
  f_\text{F}=\frac{1}{1+e^{(r-c)/a}}\ ,
\end{equation}
where $c$ refers to the half-density radius and $a$ ($a=0.52\,$fm
\cite{Piekarewicz:2016vbn}) to the surface diffuseness. The Fourier
transform can be analytically integrated resulting in
\begin{align}
  \label{eq:Fermi-symmetrized}
  F_\text{SF}=&\frac{3}{Qc}
  \left[
  \frac{\sin(Qc)}{(Qc)^2}
  \left(\frac{\pi Q a}{\tanh(\pi Qa)}\right)
  -
  \frac{\cos(Qc)}{Qc}
  \right]
  \nonumber\\
  &\times
  \left(\frac{\pi Qa}{\sinh(\pi Qa)}\right)
  \frac{1}{1 + (\pi a/c)^2}\ .
\end{align}
In this case the half-density radius $c$ proceeds from the
point-proton and point-neutron distributions mean-square radii and the
surface diffuseness $a$
\begin{equation}
  \label{eq:half-density-radius}
  c=\sqrt{\frac{5}{3}R_X^2 - \frac{7}{3}(\pi a)^2}
  \qquad (X=p,n)\ .
\end{equation}
We now calculate the weak-charge form factor percentage difference (as
a ``measure'' of the percentage spread implied by parametrization
choice) obtained by calculating $F_\text{W}$ according to
Eq. (\ref{eq:EW-FF-approx}), using for the elastic vector proton and
neutron form factors the three different parametrizations already
discussed. The result is displayed in
Fig. \ref{fig:form-factor-parametrization-dependences}. The left graph
shows results for argon, while the right graph results for
cesium. Differences are slightly more pronounced for the latter, thus
demonstrating they are more relevant for heavy nuclides. As $Q$
increases, the Helm form factor decreases more steeply. This effect is
less pronounced for the Fourier transform of the symmetrized Fermi
distribution and even less for the Klein-Nystrand
parametrization. This means that event rates calculated with the Helm
form factor will produce slightly less events than those calculated
with the Fermi parametrization, and even less than those obtained with
the Klein-Nystrand form factor. This translates into percentage
differences of the order of $1\%$ (or below) for argon and $2\%$ (or
below) for cesium. In summary, percentage differences due to form
factor parametrizations of the elastic vector proton and neutron form
factors are comparable to those implied by the $Q$-dependent nucleon
form factor terms discussed in the previous Section.
\subsection{Model-independent versus 
  form factor parametrizations approaches}
\label{sec:model-ind-vs-model-dep}
Assuming the nucleon distributions to be spherically symmetric
(i.e. assuming that the charge density distribution depends solely on
the distribution radius) leads to the series expansions of the elastic
vector proton and neutron form factors,
Eqs. (\ref{eq:FF-nuclear-expansion-p}) and
(\ref{eq:FF-nuclear-expansion-n}). These expansions, subject only to
the spherical symmetry hypothesis, involve the point-nucleon
mean-square radii (second radial moment) at order $Q^2$. So rather
than sticking to a form factor parametrization or a nuclear physics
model one can use these expansions to fit $R_n^2$ to data
\cite{Patton:2012jr,Papoulias:2019lfi}. The question is of course
whether a $Q^2$-order description suffices, or whether higher order
terms should be included to increase
convergence. Ref. \cite{Patton:2012jr} addressed this question and
ended up concluding that order $Q^4$ terms are required (for argon for
which the analysis was done). Following this approach, this implies
the introduction of a new parameter, the fourth radial moment of the
nucleon distributions, $\langle R^4_X\rangle$ ($X=p,n$) \footnote{Note
  that we have simplified our notation for the point-nucleon
  mean-square radii $R_X^2\equiv\langle R^2_X\rangle$.  For
  $\langle R^4_X\rangle$ we do not do so to avoid confusion.} .

\begin{figure*}
  \centering
  \includegraphics[scale=0.45]{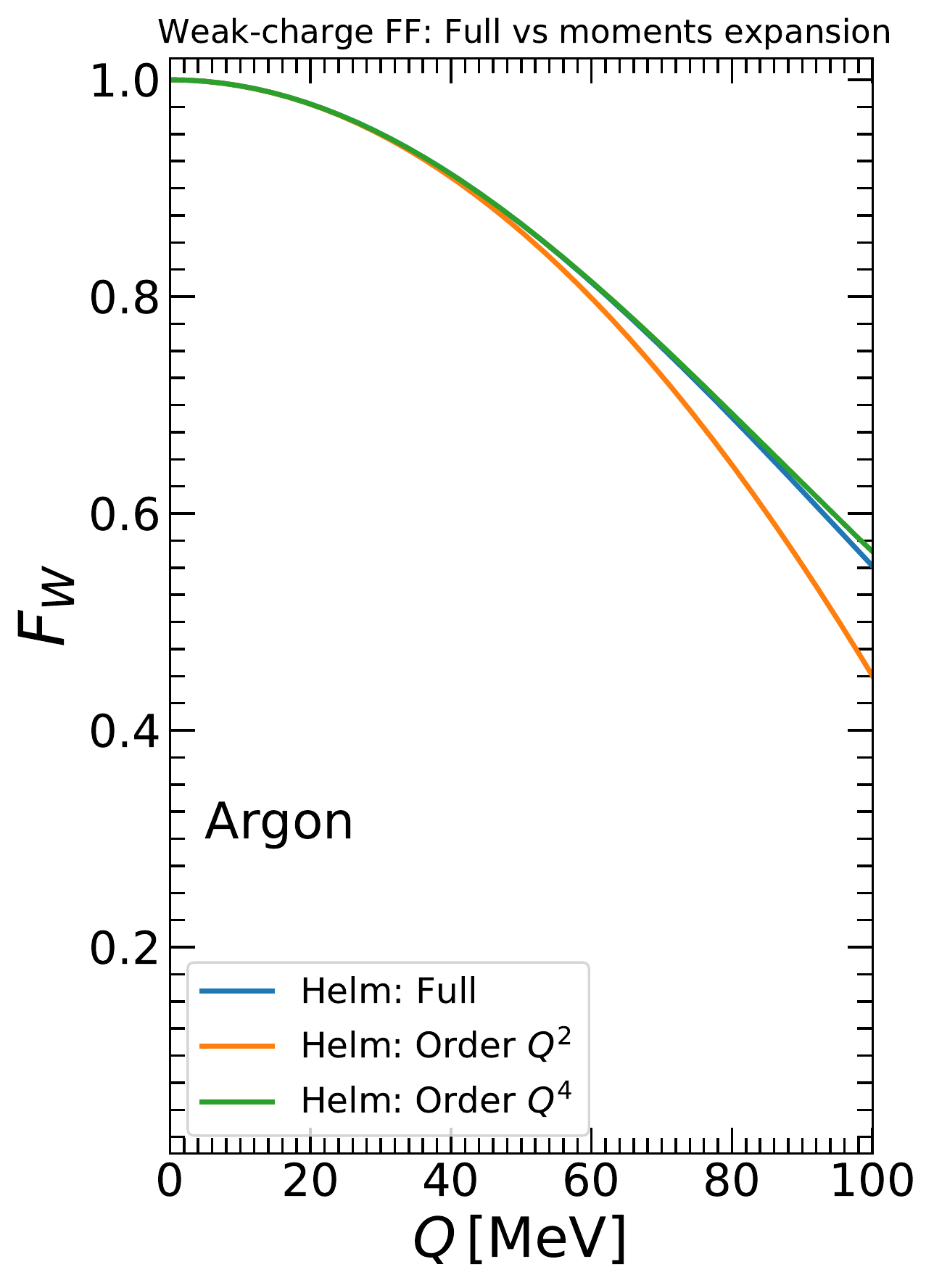}
  \includegraphics[scale=0.45]{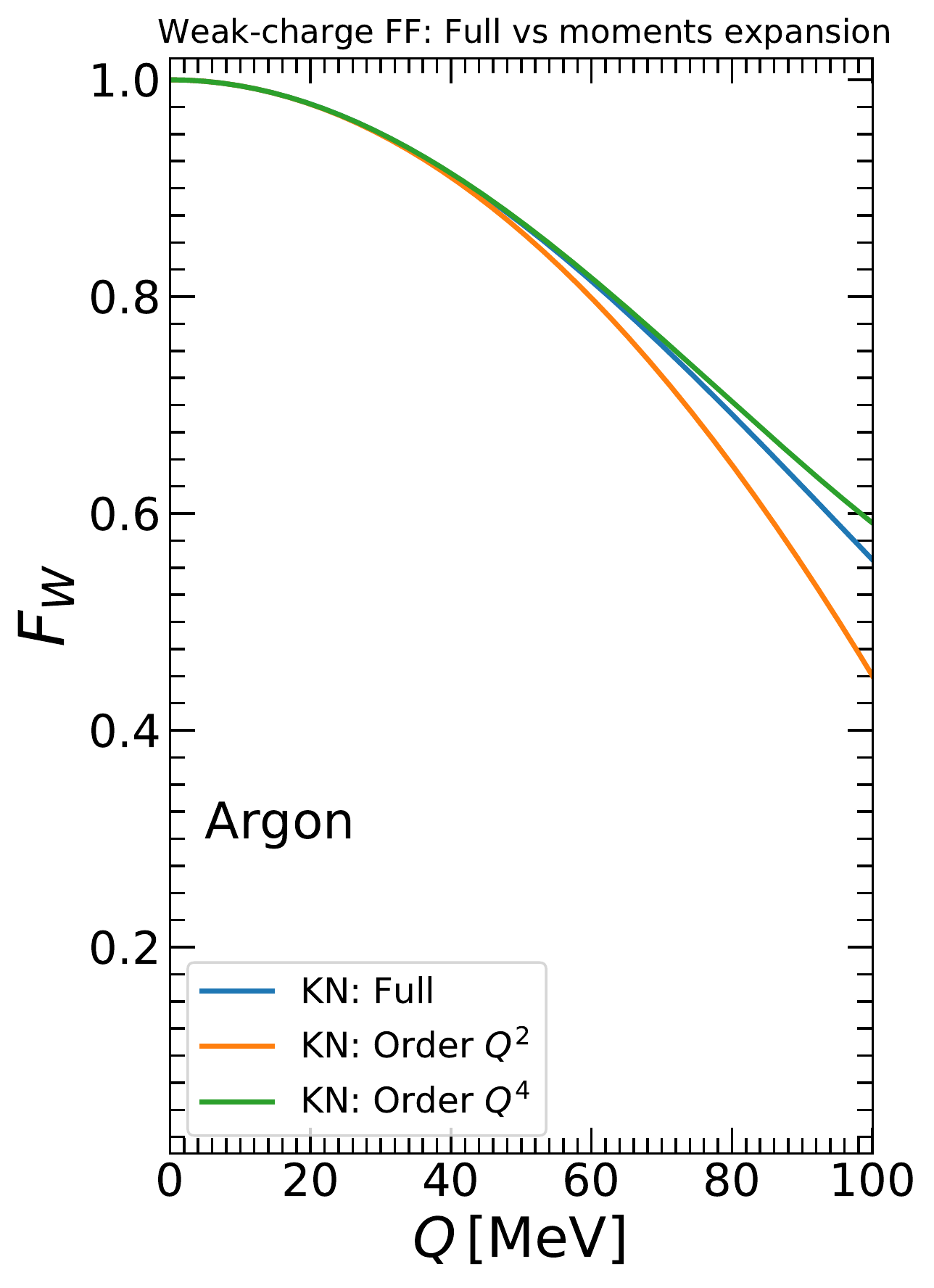}
  \includegraphics[scale=0.45]{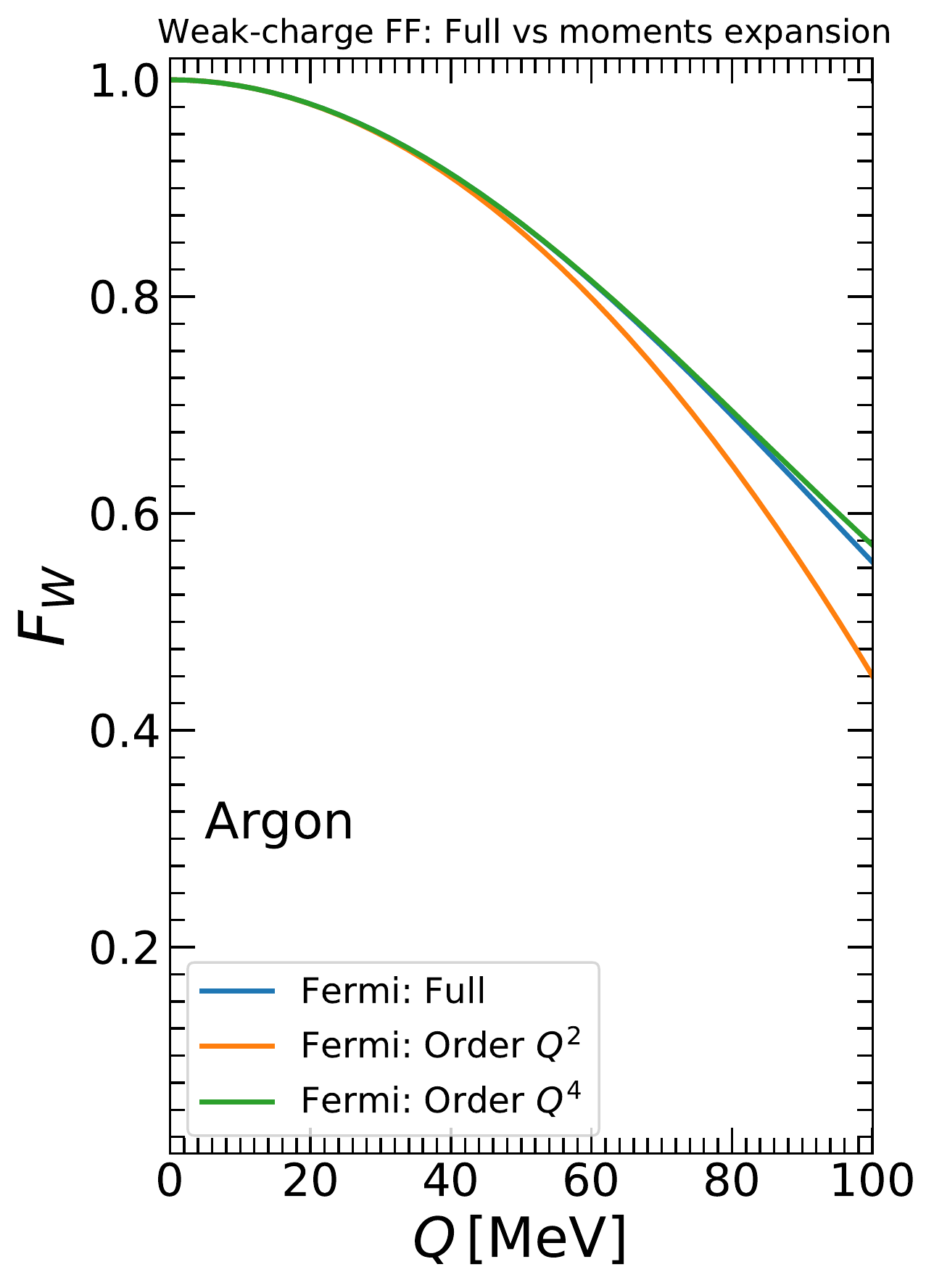}
  \includegraphics[scale=0.45]{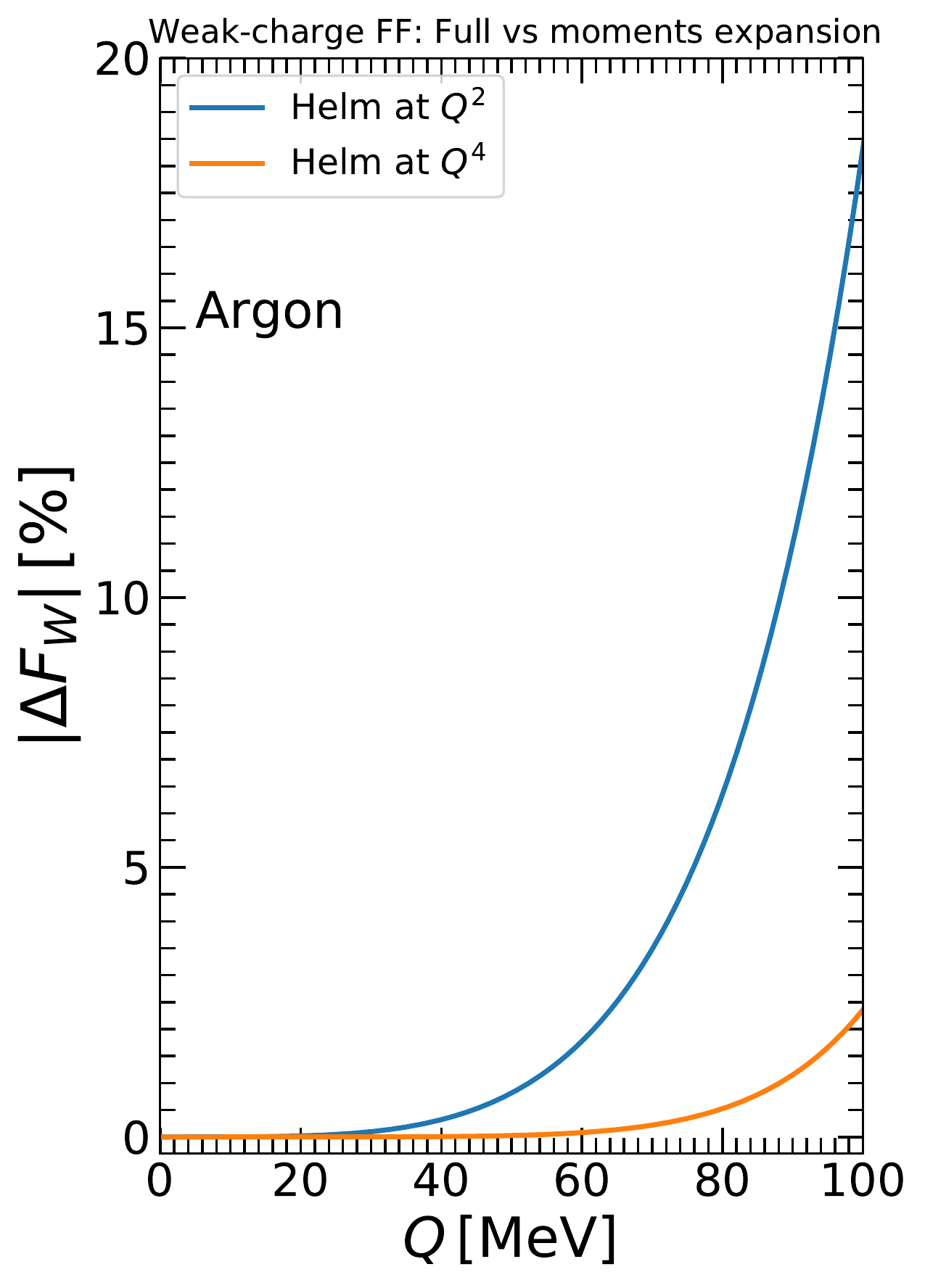}
  \includegraphics[scale=0.45]{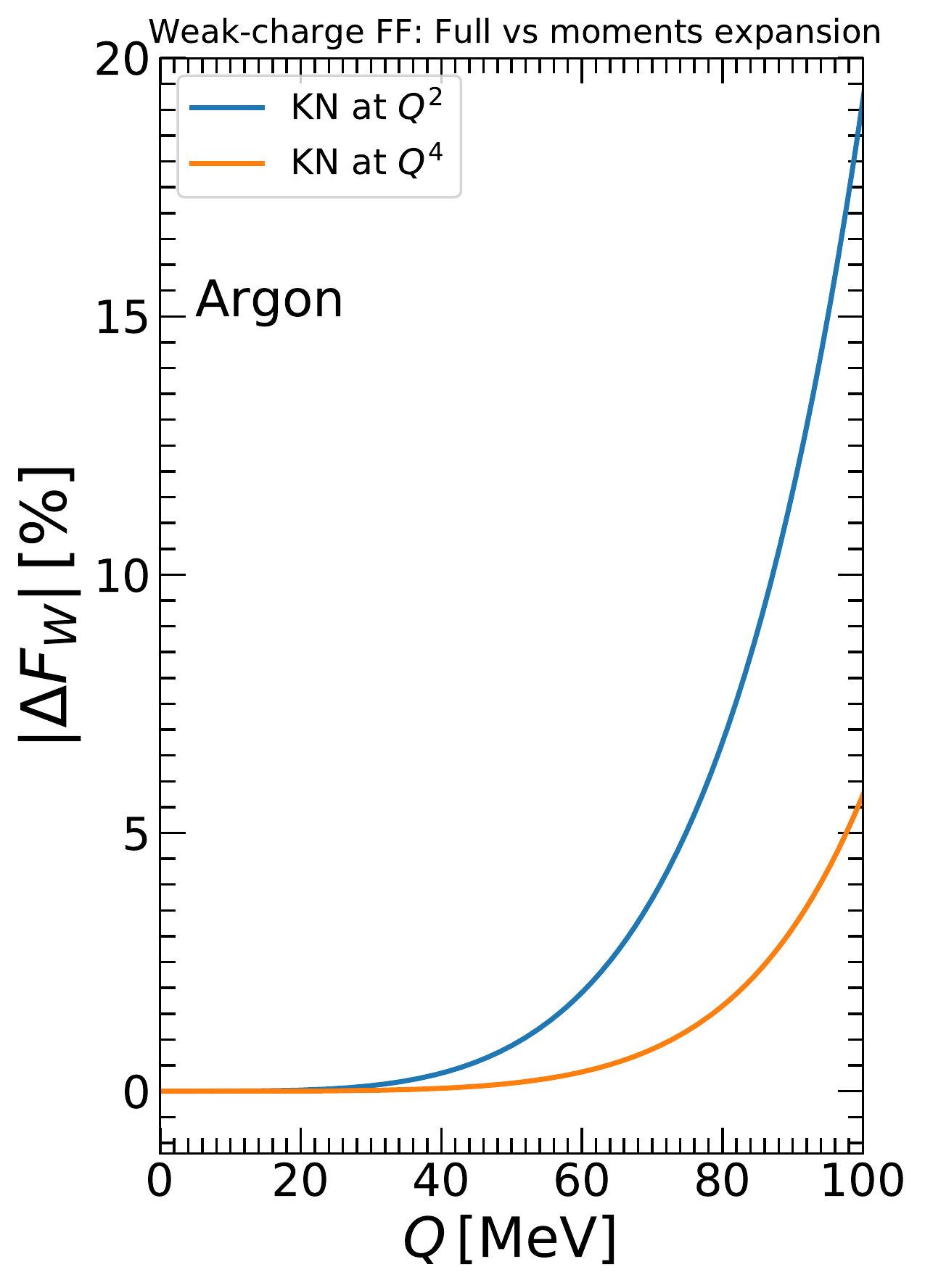}
  \includegraphics[scale=0.45]{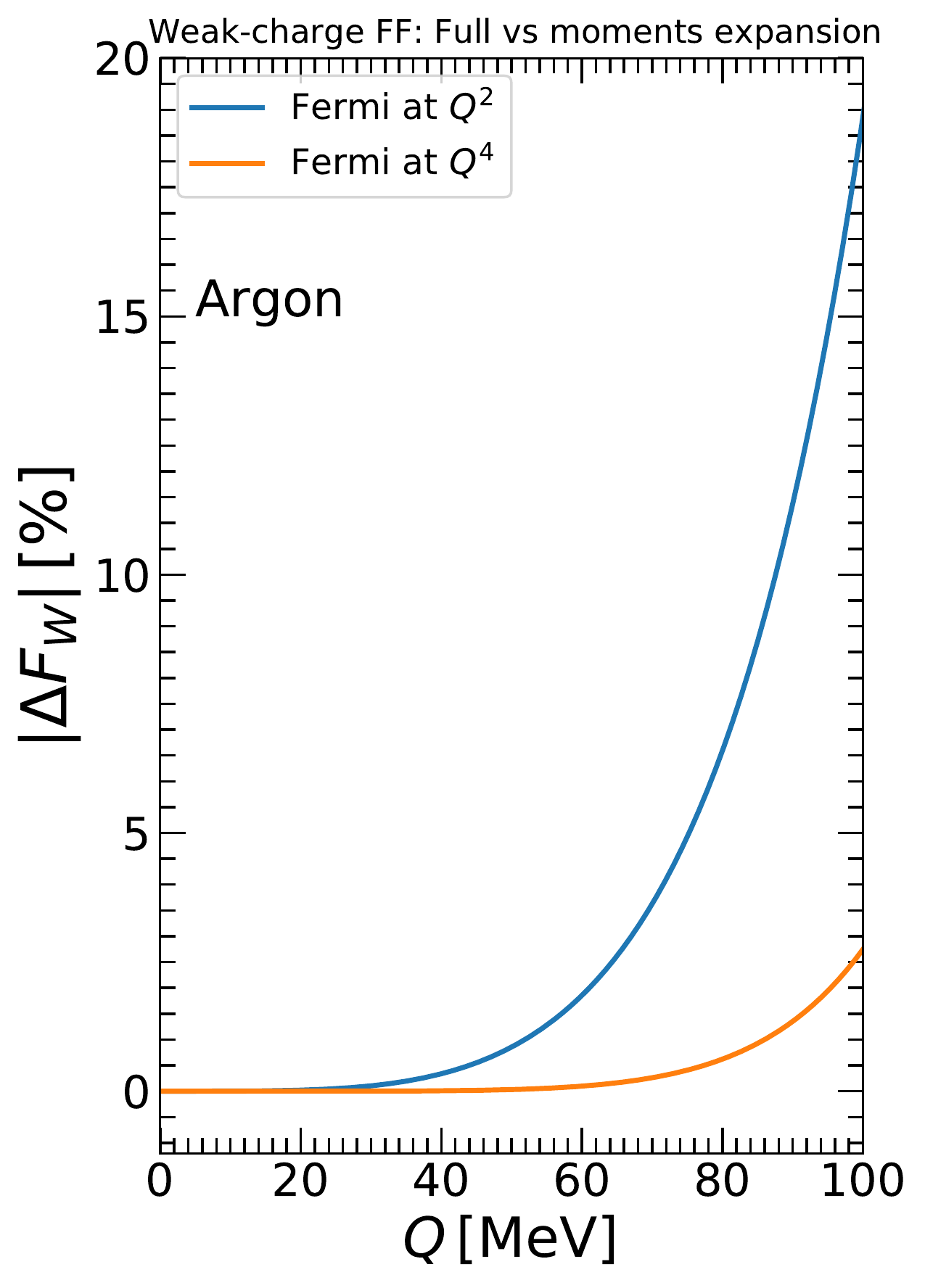}
  \caption{\textbf{Top graphs}: Convergence level for the weak-charge
    form factor series expansions at order $Q^2$, $Q^4$ and $Q^6$ (in
    argon) for the Helm, Klein-Nystrand and the Fourier transform of
    the symmetrized Fermi distribution. Expansion at order $Q^n$
    involve radial moments up to $n$-th order (see
    text). \textbf{Bottom graphs}: Percentage uncertainty in each
    case. From these results one can see that if one relies on elastic
    vector form factor power expansions and demands precision below a
    few percent the fourth radial moment should be included for light
    nuclides. Inline with Ref. \cite{Patton:2012jr}. For the $Q$ range
    displayed in the graphs the neutrino flux is sizable (see text for
    more details).}
  \label{fig:model-independent-versus-model-dependent-Ar}
\end{figure*}

\begin{figure*}
  \centering
  \includegraphics[scale=0.45]{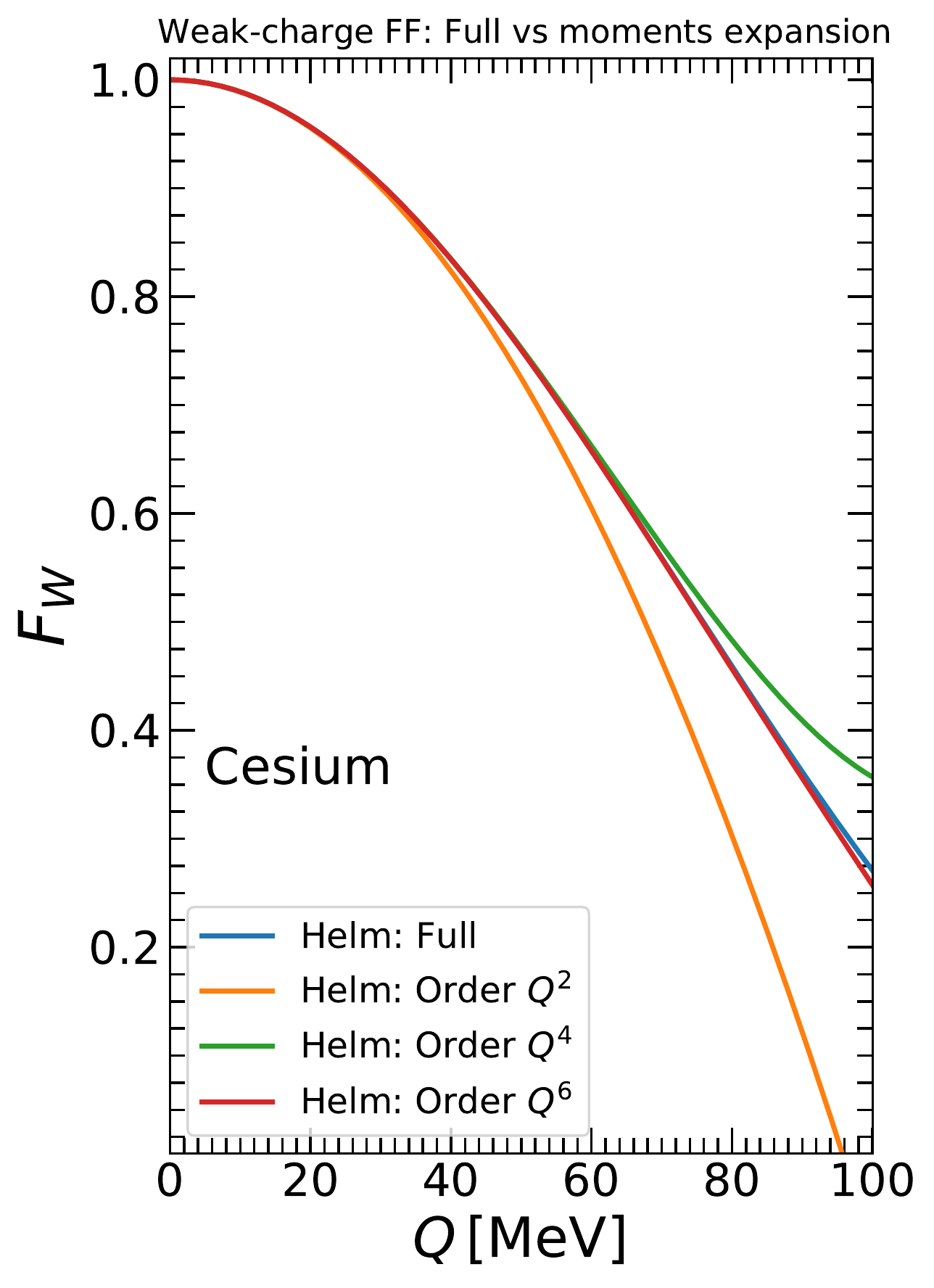}
  \includegraphics[scale=0.45]{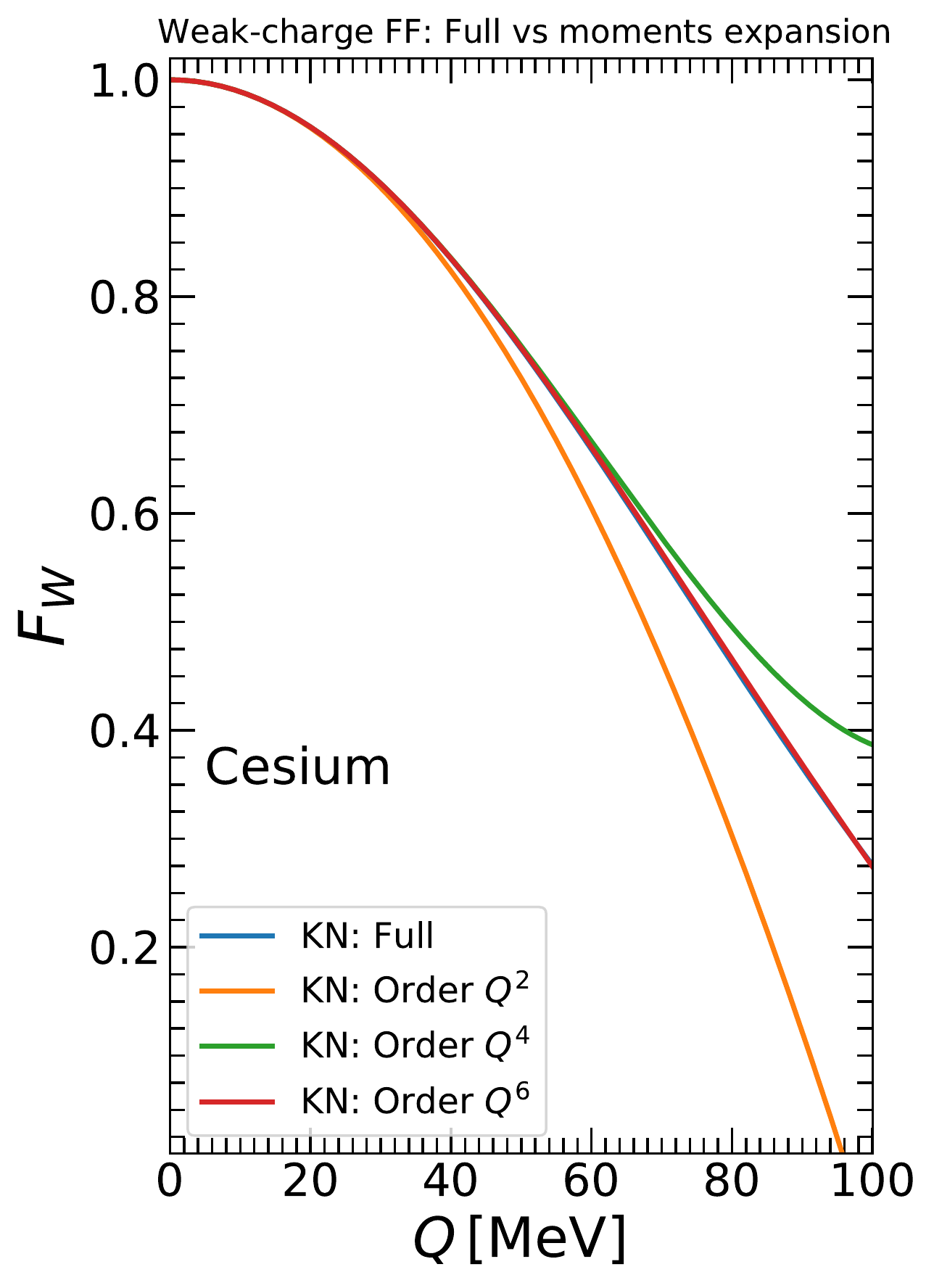}
  \includegraphics[scale=0.45]{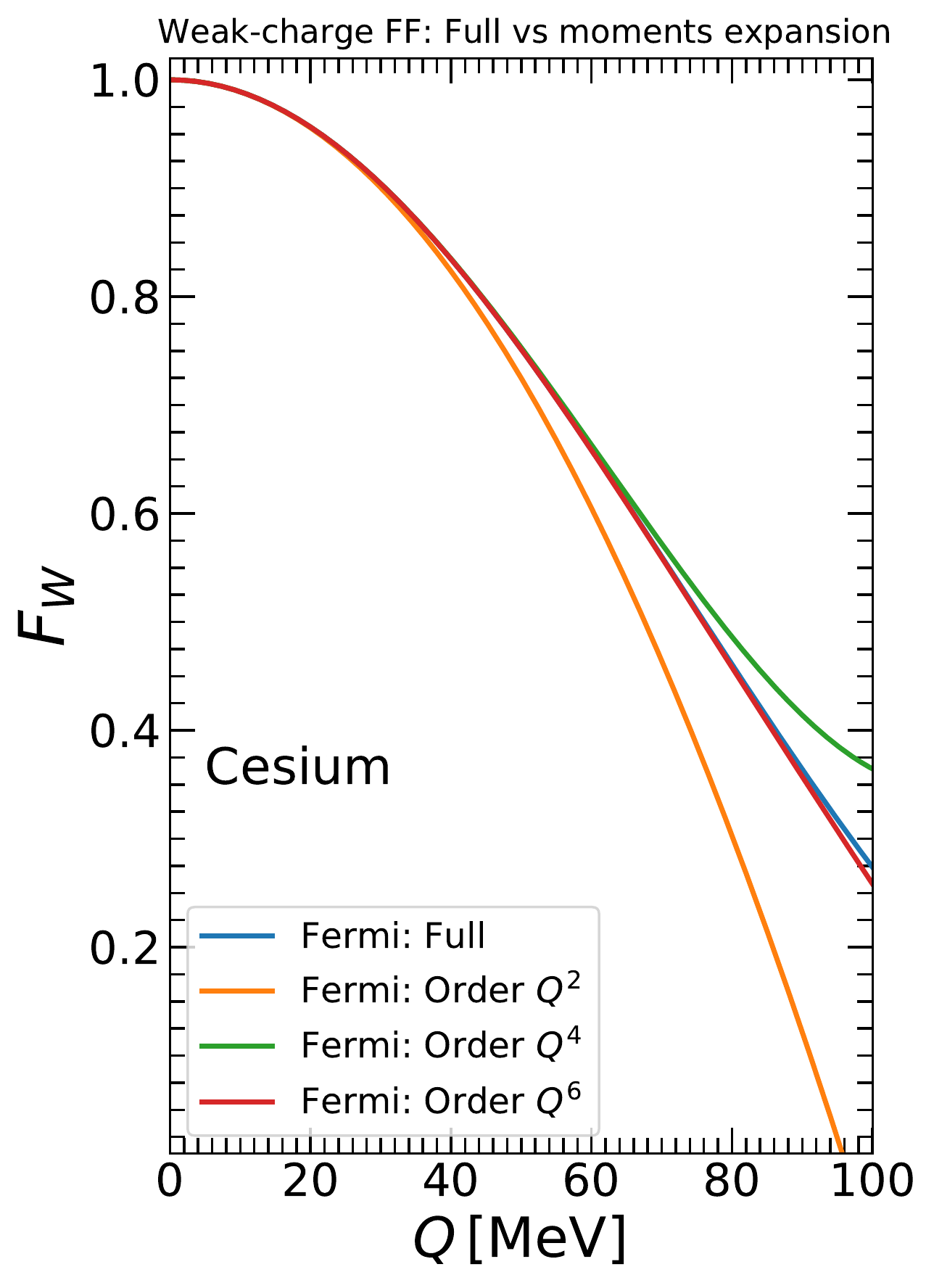}
  \includegraphics[scale=0.45]{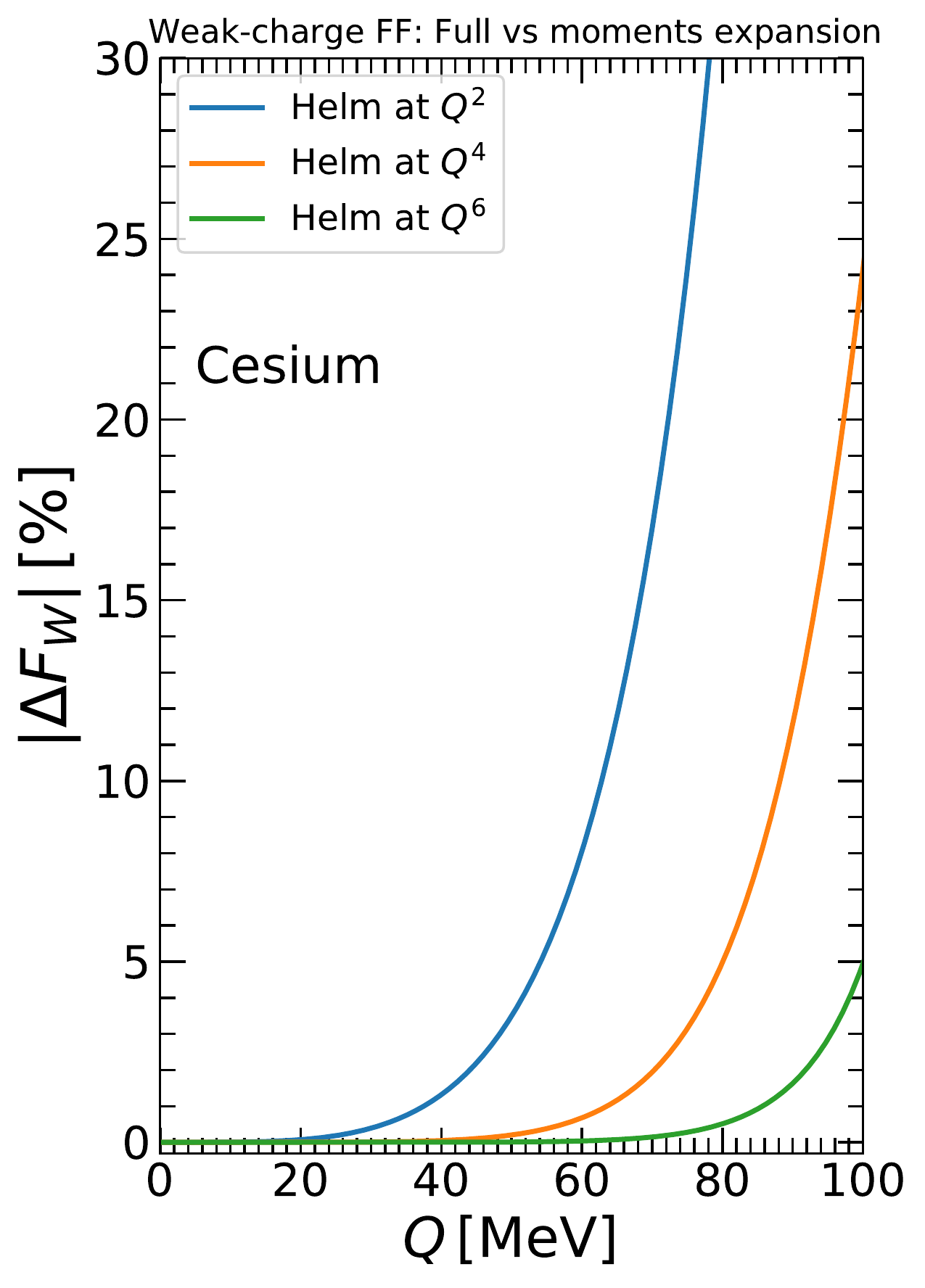}
  \includegraphics[scale=0.45]{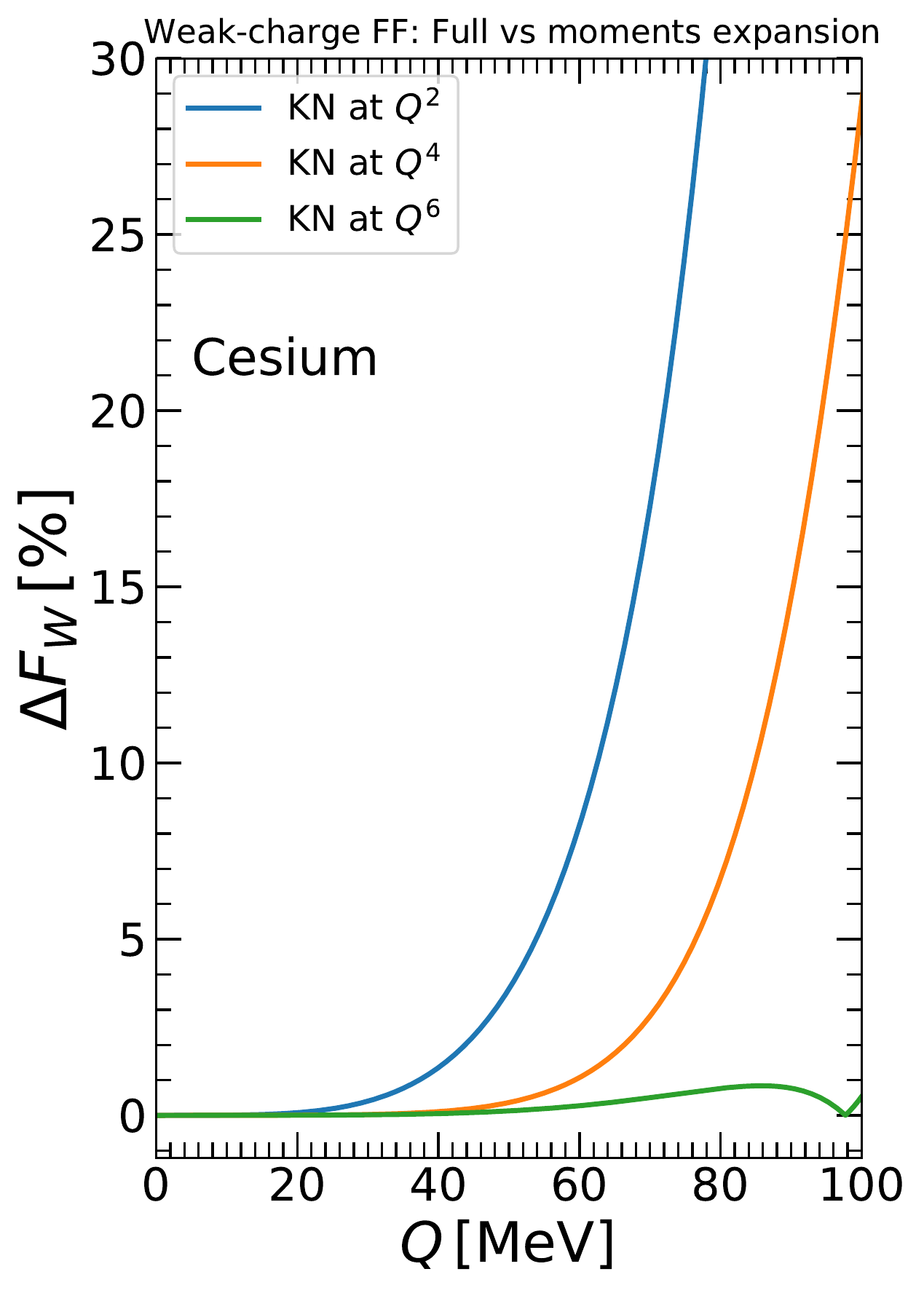}
  \includegraphics[scale=0.45]{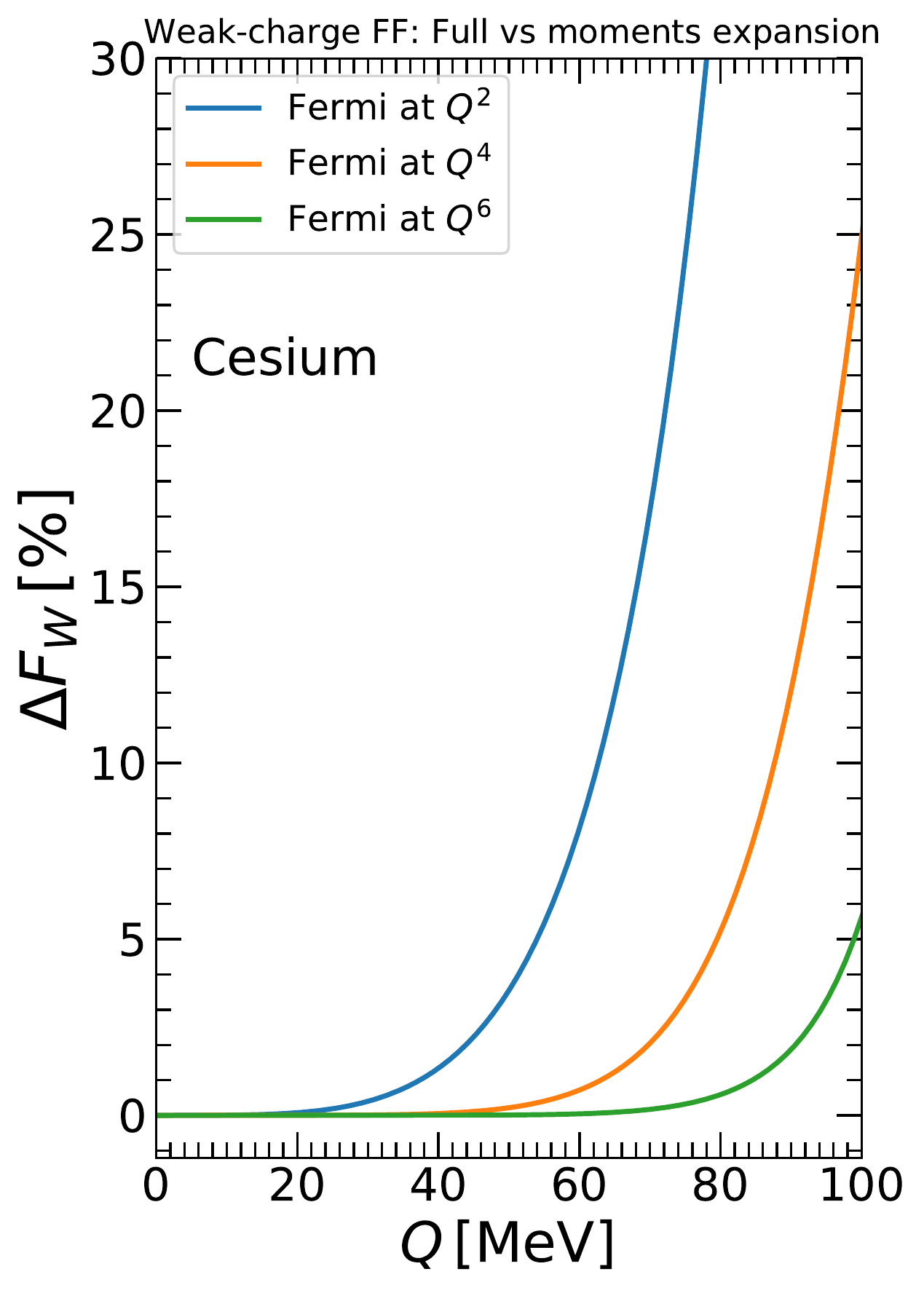}
  \caption{\textbf{Top graphs}: Convergence level for the weak-charge
    form factor series expansions at order $Q^2$, $Q^4$ and $Q^6$ (in
    cesium) for the Helm, Klein-Nystrand and the Fourier transform of
    the symmetrized Fermi distribution. Expansion at order $Q^n$
    involve radial moments up to $n$-th order (see
    text). \textbf{Bottom graphs}: Percentage uncertainty in each
    case. From these results one can see that if one relies on elastic
    vector form factor power expansions and demands precision below a
    few percent the sixth radial moment should be included. Inline as
    well with Ref. \cite{Patton:2012jr}. For the $Q$ range displayed
    in the graphs the neutrino flux is sizable (see text for more
    details).}
  \label{fig:model-independent-versus-model-dependent-Cs}
\end{figure*}

To compare the accuracy of the model-independent analysis and the
form factor parametrization approach, we first calculate the fourth
radial moment for the three parametrizations we are using. We do so by
taking into account that the series expansions in
Eqs. (\ref{eq:FF-nuclear-expansion-p}) and
(\ref{eq:FF-nuclear-expansion-n}) can be compared term by term to the
Taylor expansions of the $F_V^p(Q^2)$ and $F_V^n(Q^2)$ functions. Up to
order $Q^4$ this reduces to
\begin{equation}
  \label{eq:derivates_and_moments}
  R^2_X = 
  -6\left . 
    \frac{d F_V^X}{d Q^2}
  \right|_{Q^2=0}
  \quad\text{and}\quad
  \langle R^4_X\rangle =
  60 \left . 
    \frac{d^2 F_V^X}{d Q^4}
  \right|_{Q^2=0}\ .
\end{equation}
More generally, the $2i$-th moment can be written according to
\begin{equation}
  \label{eq:2ith-moment}
  \langle R_X^{2i}\rangle = (-1)^i\frac{(2i+1)!}{i!}
  \left .\frac{d^iF_V^X}{dQ^{2i}}\right|_{Q^2=0}\quad (X=p,n)\ .
\end{equation}
The first relation in Eq. (\ref{eq:derivates_and_moments}) leads to
the expressions for the diffraction, hard sphere and half density
radii ($R_0$, $R_A$ and $c$) for the Helm, Klein-Nystrand and Fourier
transform of the symmetrized Fermi distribution,
Eqs. (\ref{eq:diffraction-radius-mean-square-radius}),
(\ref{eq:radius-potential-range-KN-FF}) and
(\ref{eq:half-density-radius}). The second relation in
Eq. (\ref{eq:derivates_and_moments}) leads instead to
\begin{alignat}{2}
  \label{eq:four-moment-H-KN-F}
  &\text{Helm:}&\quad\langle R^4_X\rangle=&\frac{3}{7}R_0^4+6R_0^2s^2+15s^4\ ,
  \nonumber\\
  &\text{KN:}&\quad\langle R^4_X\rangle=&\frac{3}{7}R_A^4+12a_k^2R_A^2+120a_k^4\ ,
  \nonumber\\
  &\text{Fermi:}&\quad\langle R^4_X\rangle=&\frac{3}{7}c^4 
  + \frac{18}{7} c^2(a\pi)^2 + \frac{31}{7}(a\pi)^4\ .
\end{alignat}
Of course the higher the order in $Q$ the expansion extends to, the
better the convergence and so the reliability of the result. Inclusion
of terms up to order $Q^6$ require the sixth radial moment. From
Eq. (\ref{eq:2ith-moment}) explicit expressions for each case read
\begin{alignat}{2}
  \label{eq:sixth-moment-H-KN-F}
  &\text{Helm:}&\,\langle R^6_X\rangle=&\frac{1}{3}R_0^6+9R_0^2s^2+63R_0^2s^4+105 s^6\ ,
  \nonumber\\
  &\text{KN:}&\,\langle R^6_X\rangle=&\frac{1}{3}R_A^6+18R_A^4a_k^2+504R_k^2a_k^4+5040a_k^6\ ,
  \nonumber\\
  &\text{Fermi:}&\,\langle R^6_X\rangle=&\frac{1}{3}c^6 + \frac{11}{3} c^4(\pi a)^2 
  + \frac{239}{15}c^2(\pi a)^4+\frac{127}{5}(\pi a)^6\ .
\end{alignat}
Using Eqs. (\ref{eq:diffraction-radius-mean-square-radius}),
(\ref{eq:radius-potential-range-KN-FF}) and
(\ref{eq:half-density-radius}) for $R_0$, $R_A$ and $c$ one can write
the fourth and sixth radial moments for each parametrization solely in
terms of $R_X^2$. In doing so one can then compare the level of
convergence of the $Q^2$, $Q^4$ and $Q^6$ expansions with the full
expressions for each form factor. Results are shown in
Figs. \ref{fig:model-independent-versus-model-dependent-Ar} and
\ref{fig:model-independent-versus-model-dependent-Cs} for argon and
cesium, respectively (representative of light and heavy nuclides).

Top graphs in
Fig. \ref{fig:model-independent-versus-model-dependent-Ar} show
results for the three form factor parametrizations calculated for
argon. Bottom graphs show percentage uncertainties for each case. One
can see that inclusion of only the second moment leads to deviations
above $\sim 5\%$ for transferred momenta of the order of $80\,$MeV. At
$100\,$MeV those deviations can reach $\sim 20\%$. These
uncertainties, however, should be taken with care as those found in
the previous Section for the same reason. They increase at relatively
large $Q$, where eventually the neutrino flux fades away. Thus, we
have plotted up to $Q$ values where still an abundant number of
neutrinos is available. They will---of course---have an impact when
comparing results from both approaches, but the best way to understand
their actual impact is through the calculation of the CE$\nu$NS event
rate, something that will be done and discussed in
Sec. \ref{sec:neutron-distribution-limits}.

Top and bottom graphs in
Fig. \ref{fig:model-independent-versus-model-dependent-Cs}
show---instead---results for cesium. In this case a few percentage
convergence requires the inclusion of the sixth moment. As can be seen
in the bottom graphs, up to order $Q^4$ the expansion involves
uncertainties that can readily reach $\sim 10\%$ at transferred
momenta of the order of $80\,$MeV, values for which the neutrino flux
is still sizable enough for the uncertainty to have an impact on the
CE$\nu$NS event rate. Aiming at few percent level precision thus
requires the inclusion of the sixth moment, inline with
Ref. \cite{Patton:2012jr}.

The adoption of form factor parametrizations suffers from the model
dependence implied by the different assumptions those parametrizations
come along with. However, as we have discussed in
Sec. \ref{sec:form_factor_param}, using one or the other (which can be
understood as a variation of the underlying nuclear physics
hypotheses) introduces uncertainties at the few percent level. The
same level of precision can be achieved with the model-independent
power expansion approach, which relies only on the assumption of a
spherical symmetric nuclear ground state. However to achieve that
level of precision new parameters should be considered. Thus, take for
instance the case of cesium. Given a CE$\nu$NS dataset an analysis
relying on form factor parametrizations is expected to imply a few
percent uncertainty (in addition to systematic and statistical
uncertainties due to e.g. quenching factor, neutrino flux, BRN and SS
backgrounds). An analysis based on radial moments expansions as well,
but then the dataset has to be used to fit not only the neutron
distribution rms radius but also the fourth and sixth moments
(depending on the nuclide the detector is built with). The extraction
procedure then becomes a multiparameter problem, which might worsen
the precision with which $R_n^2$ can be determined. We will come back
to that discussion in Sec. \ref{sec:neutron-distribution-limits}.
\section{CE$\nu$NS cross section and event rate}
\label{eq:cevns-xsec}
The CE$\nu$NS differential cross section follows from a neutral
current process. In terms of nuclear recoil energy it is given by
\cite{Stodolsky:1966zz,Freedman:1973yd,Kopeliovich:1974mv,Freedman:1977xn}
\begin{equation}
  \label{eq:CEvNS_xsec}
  \frac{d\sigma}{dE_r}=\frac{G_F^2 m_N}{2\pi}\,Q_W^2\,F_W^2
  \left(2 - \frac{m_N E_r}{E_\nu^2}\right)\ ,
\end{equation}
where subleading kinematic terms have been neglected and $m_N$ here
refers to nuclear mass. The strength at which the $Z$ boson couples to
the nucleus is determined by the $Q$-dependent coupling
$Q_W\times F_W(Q^2)$. The coupling is such that with increasing
transferred momentum (increasing incoming neutrino energy) the
``effective'' weak charge decreases and so the interaction
probability. Uncertainties in $F_W$, as those we have discussed in the
previous Sections and as those discussed in
Ref. \cite{AristizabalSierra:2019zmy} \footnote{Uncertainties on the
  weak-charge form factor using the large-scale nuclear shell model
  for a long list of nuclei of interest have been discussed in
  Ref. \cite{Hoferichter:2020osn}.}, translate into uncertainties in
the CE$\nu$NS cross section. They are indeed responsible for the
nuclear physics uncertainties the process involves and so are entirely
responsible for the theoretical uncertainties of the CE$\nu$NS event
rate, unless one-loop electroweak corrections are accounted for
\cite{Tomalak:2020zfh}.

\begin{figure*}
  \centering
  \includegraphics[scale=0.395]{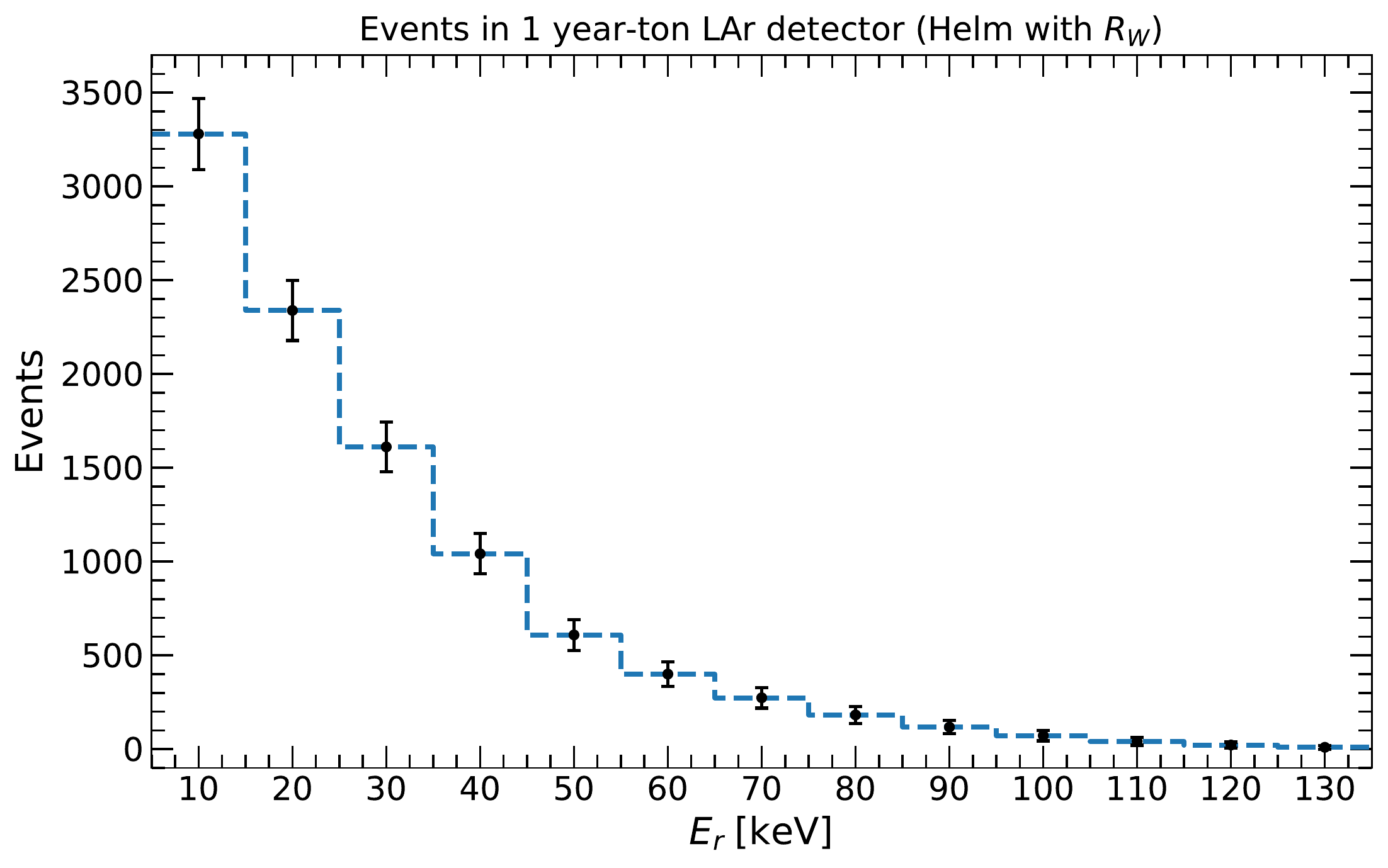}
  \includegraphics[scale=0.395]{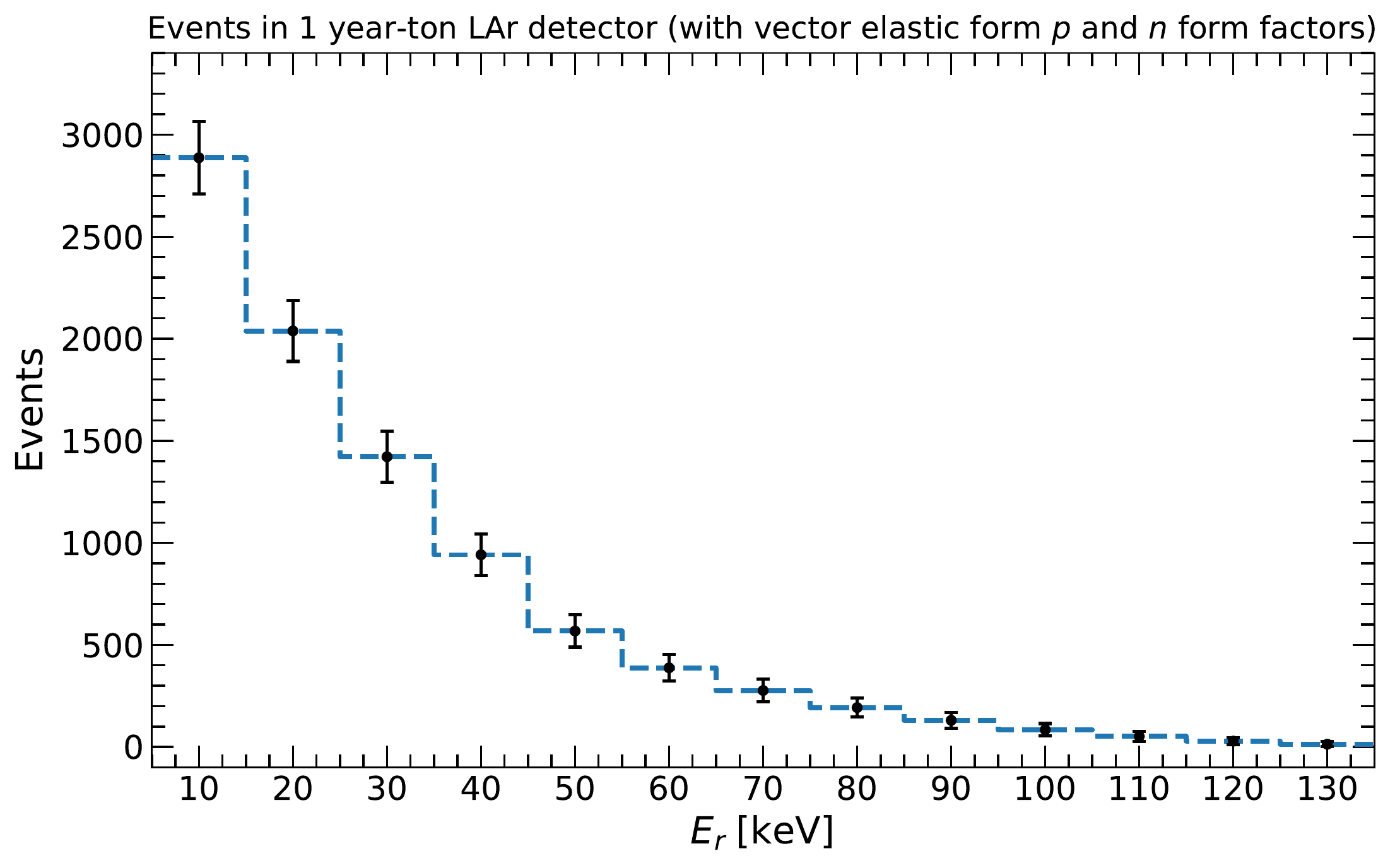}
  \includegraphics[scale=0.392]{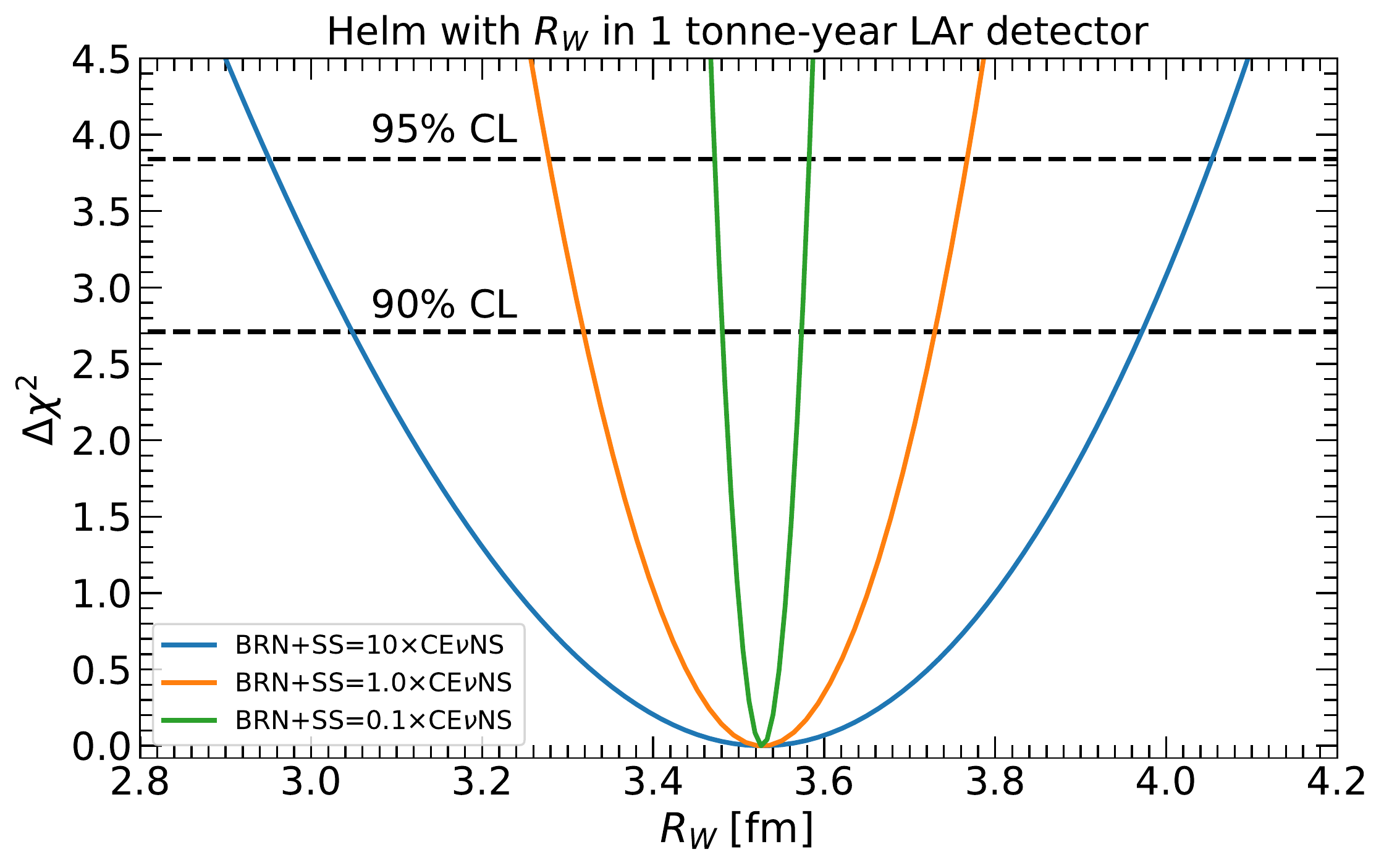}
  \includegraphics[scale=0.392]{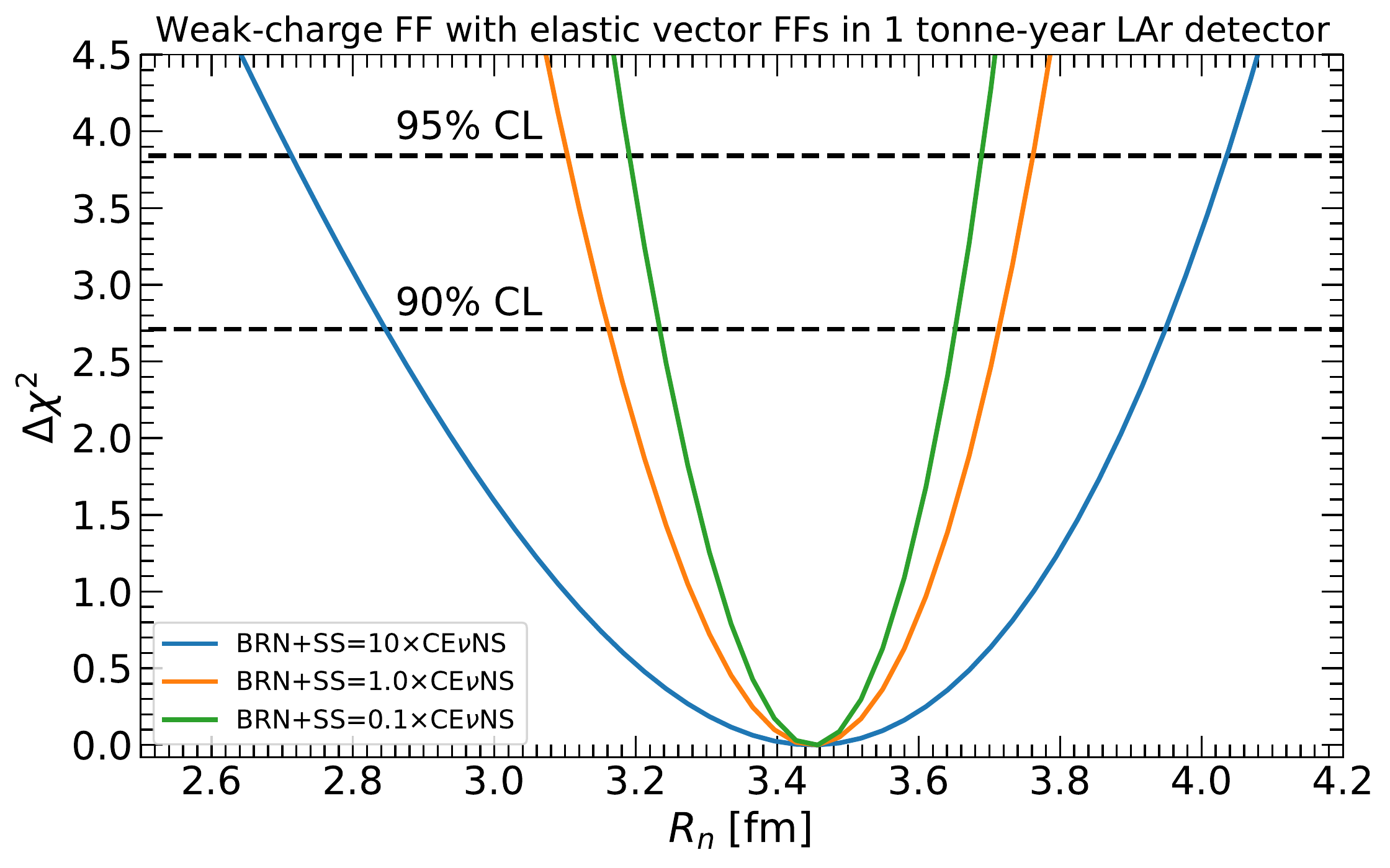}
  \caption{\textbf{Top graphs}: Toy experiment signals calculated by:
    (i) Using the Helm form factor and fixing the diffraction radius
    with $R_W$ (left graph), (ii) Using the weak-charge form factor as
    given in Eq. (\ref{eq:EW-FF-approx}) (right graph). In both cases
    $R_n=R_p + 0.1\,\text{fm}$, with $R_p$ calculated from the nuclear
    charge radius taken from Ref. \cite{Angeli:2013epw} and
    Eq. (\ref{eq:EM-EW-charge-radii}). \textbf{Bottom graphs}:
    Least-square function as a function of the weak-charge radius
    $R_W$ (left graph) and the point-neutron distribution rms radius
    (right graph). In both graphs results for two additional analyses
    with two different BRN and SS background hypotheses are shown as
    well.}
  \label{fig:toy_experiments_signals}
\end{figure*}
The CE$\nu$NS differential event rate, as any other rate, follows from
a convolution of the differential cross section in
Eq. (\ref{eq:CEvNS_xsec}) and the incoming neutrino flux. For
CE$\nu$NS to be used as a tool for the extraction of neutron
distributions mean-square radii the neutrino flux should lie in either
the ``intermediate'' or ``high'' energy windows. The former defined by
pion decay at rest, while the latter by pion decay in flight. Note
that from this perspective COHERENT experiments operate in the
intermediate energy window and the $\nu$BDX-DRIFT experiment will in
the high energy regime
\cite{AristizabalSierra:2021uob,AristizabalSierra:2022jgg}. Because of
the large statistics our analysis is interested in we focus on the
intermediate energy window, where $\mathcal{O}(\text{tonne})$
detectors are expected in the near future
\cite{ref:STS,Asaadi:2022ojm}. In this case the neutrino spectrum
consist of a monochromatic component (prompt component) and two
continuous spectra (delayed components). Their spectral functions
follow from the $\pi^+$ and $\mu^+$ energy distributions and read
\begin{align}
  \Phi_{\nu_\mu}&=\delta\left(E_\nu - \frac{m_\pi^2-m_\mu^2}{2 m_\pi}\right)\ ,
                  \nonumber\\
  \Phi_{\overline{\nu}_\mu}&=\frac{192}{m_\mu}\left(\frac{E_\nu}{m_\mu}\right)^2
                             \left(\frac{1}{2}-\frac{E_\nu}{m_\mu}\right)\ ,
                  \nonumber\\
  \Phi_{\nu_e}&=\frac{64}{m_\mu}\left(\frac{E_\nu}{m_\mu}\right)^2
                             \left(\frac{3}{4}-\frac{E_\nu}{m_\mu}\right)\ .
\end{align}
Since neutrinos are isotropically produced, normalization of the
neutrino flux is given by $n_\nu=r\times\text{POT}/4/\pi^2/L^2$. We
use $\text{POT}=2.1\times 10^{23}/\text{year}$, $L=28.0\;\text{m}$ and
$r=8.0\times 10^{-2}$, inspired by recent COHERENT measurements and
future plans
\cite{Akimov:2017ade,Akimov:2020czh,COHERENT:2021xmm,ref:STS}. Thus
taking into account the three neutrino components, the differential
recoil spectrum in a detector with a fiducial volume composed of
different stable isotopes is written as
\begin{equation}
  \label{eq:DRS}
  \frac{dR}{dE_r}=\frac{m_\text{det}N_A}{m_{\text{mol},i}}n_\nu\sum_i X_i
  \sum_{\alpha=\nu_\mu,\overline{\nu}_\mu,\nu_e}\int_{E_\nu^\text{min}}^{E_\nu^\text{max}}
  \Phi_\alpha\,\frac{d\sigma_i}{dE_r}\,E_\nu\ .
\end{equation}
Here $m_\text{det}$ refers to the detector fiducial volume mass,
$N_A=6.022\times 10^{23}/\text{mol}$ to the Avogadro number,
$m_{\text{mol},i}$ to the molar mass of the $i^\text{\underline{th}}$
isotope and $X_i$ to its relative abundance. Lower and upper
integration limits are given by $E_\nu^\text{min}=\sqrt{E_r m_N/2}$
and $E_\nu=m_\mu/2$, respectively. Assuming uniform bin size
$\Delta E_r$, the total event rate evaluated in the
$k^\text{\underline{th}}$ bin central value $E_r^k$ follows from
integration, namely
\begin{equation}
  \label{eq:tot_event_rate}
  N = \int_{E_r^k - \Delta E_r/2}^{E_r^k + \Delta E_r/2}\frac{dR}{dE_r}\,dE_r\ .
\end{equation}

With the results from Secs. \ref{sec:charge-and-weak-FFs} and
\ref{sec:quantitative-differences}, along with those discussed in this
section, we are now in a position to study the precision with which
the point-neutron distribution rms can be extracted from data under
different weak-charge form factor approaches. For that aim we consider
a one-tonne LAr detector as we now discuss.
\section{Extraction of the argon neutron distribution root-mean-square
  radius using different approaches}
\label{sec:neutron-distribution-limits}
We now proceed with the determination of the neutron distribution rms
radius in some of the different approaches we have discussed. We focus
on three cases: \textbf{(i)} The weak-charge form factor parametrized
\`a la Helm (for definitiveness), \textbf{(ii)} the weak-charge form
factor written as in Eq. (\ref{eq:EW-FF-approx}) with the elastic
vector form factors parametrized as well \`a la Helm, \textbf{(iii)}
model-independent approach based on expansions in terms of
even-moments as discussed in
Sec.~\ref{sec:model-ind-vs-model-dep}. Note that any beyond-the-SM
physics effect will diminish the precision with which the neutron
distribution rms radius can be extracted. In the following analyses,
therefore, those effects are ignored.

Rather than using existing CsI and LAr data
\cite{Akimov:2017ade,COHERENT:2021xmm,Akimov:2020czh} we instead
assume a one-tonne LAr detector with specifications as explained in
Sec. \ref{eq:cevns-xsec}. This choice provides flexibility with
background and so enable us to highlight the main differences arising
in each case. As we have already stressed plans for deploying an
$\mathcal{O}(\text{tonne})$-scale LAr detector at the STS
\cite{ref:STS} have been discussed in Ref. \cite{Asaadi:2022ojm}.

To proceed we define a simplistic spectral least-square function that
accounts only for energy spectral data, but encapsulates the main
features of such an experimental environment: Uncertainties from
quenching factor, neutrino flux and efficiency as well as Beam Related
Neutron (BRN) and Steady State (SS) backgrounds. We, however, do not
consider systematic uncertainties due to energy calibration and
pulse-shape discrimination and assume that the neutrino-induced
neutron (NIN) background is subdominant compared with the BRN and SS
backgrounds. Note that such assumptions do not aim whatsoever at
representing the actual experimental setup, they are just choices that
enable pointing out the effects we are aiming at. The spectral
least-square function then reads
\begin{equation}
  \label{eq:least-square}
  \chi^2 = 
  \sum_{i=1}^5
    \left[\frac{N^\text{Exp}_i - (1+\alpha)N^\text{Th}_i(\mathcal{P})}
      {\sigma_i}\right]^2 + 
    \left(\frac{\alpha}{\sigma_\alpha}\right)^2\ .
\end{equation}
Here $N^\text{Exp}_i$ refers to the number of events in the
$i^\text{\underline{th}}$ bin generated in a toy experiment and
$\alpha$ a nuisance parameter. The least-square function becomes a
function of $\alpha$, with its actual value following from
minimization over it. $N^\text{Th}_i$ the number of events generated
by varying the set $\mathcal{P}=\{R_n^2,\langle R_n^4\rangle\}$ over a
certain range, where $\langle R_n^4\rangle$ is only present in case
\textbf{(ii)}. For the statistical uncertainty we assume
$\sigma_i^2=N_i^\text{Th}+\sum_j \text{B}_j$, with $\sum_j \text{B}_j$
the BRN and SS related backgrounds which we assume to follow the same
energy dependence of the signal. The latter assumption is certainly
not the case as can be seen in the LAr CE$\nu$NS measurement release
\cite{Akimov:2020czh}.  However, rather than extrapolating that
background to the case we are interested in (multi-ton LAr detector)
we believe this assumption represents a better choice. At present
$\text{BR}_\text{CE$\nu$NS}\simeq 3\%$ while
$\text{BR}_\text{BRN}\simeq 14\%$ and
$\text{BR}_\text{SS}\simeq 83\%$, but reduction of that background to
lower levels seems achievable in the future \cite{Akimov:2018ghi}. We
then assume $\text{B}_j=10\times N^\text{Exp}_j$ \footnote{Note that
  at present BRN plus SS backgrounds in the LAr data release amount to
  about 29 times the CE$\nu$NS signal \cite{Akimov:2020czh}.}. For the
systematical uncertainty encoded in $\sigma_\alpha$ we take
$\sigma_\alpha=0.11$ from a $10\%$ uncertainty due to neutrino flux,
$5\%$ uncertainty due to efficiency and $1\%$ due to quenching
factor. For the efficiency, and following once again the CENNS-10 LAr
detector, we assume a Heaviside function at $5\,\text{keV}_\text{nr}$
and a $10\text{keV}_\text{nr}$ bin size. To generate the toy
experiment data, we fix the point-proton distribution rms radius with
the aid of Eq. (\ref{eq:EM-EW-charge-radii}) with the values for the
different quantities given in
Tab. \ref{tab:argon-germanium-parameters}. For the point-neutron
distribution rms radius we use $R_n=R_p+0.1\,\text{fm}$, as we have
done in the previous Sections.

\begin{figure*}
  \centering
  \includegraphics[scale=0.397]{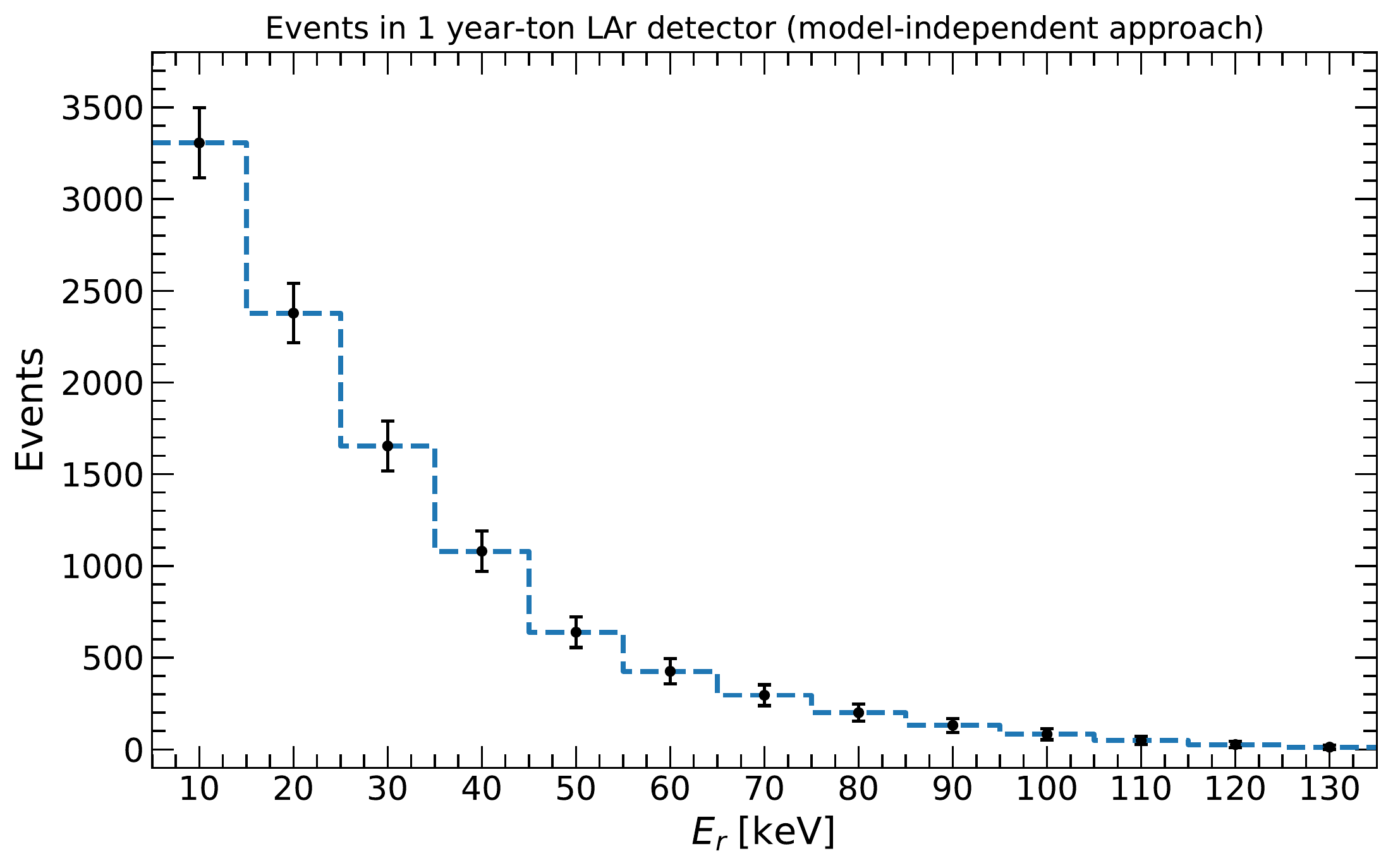}
  \includegraphics[scale=0.397]{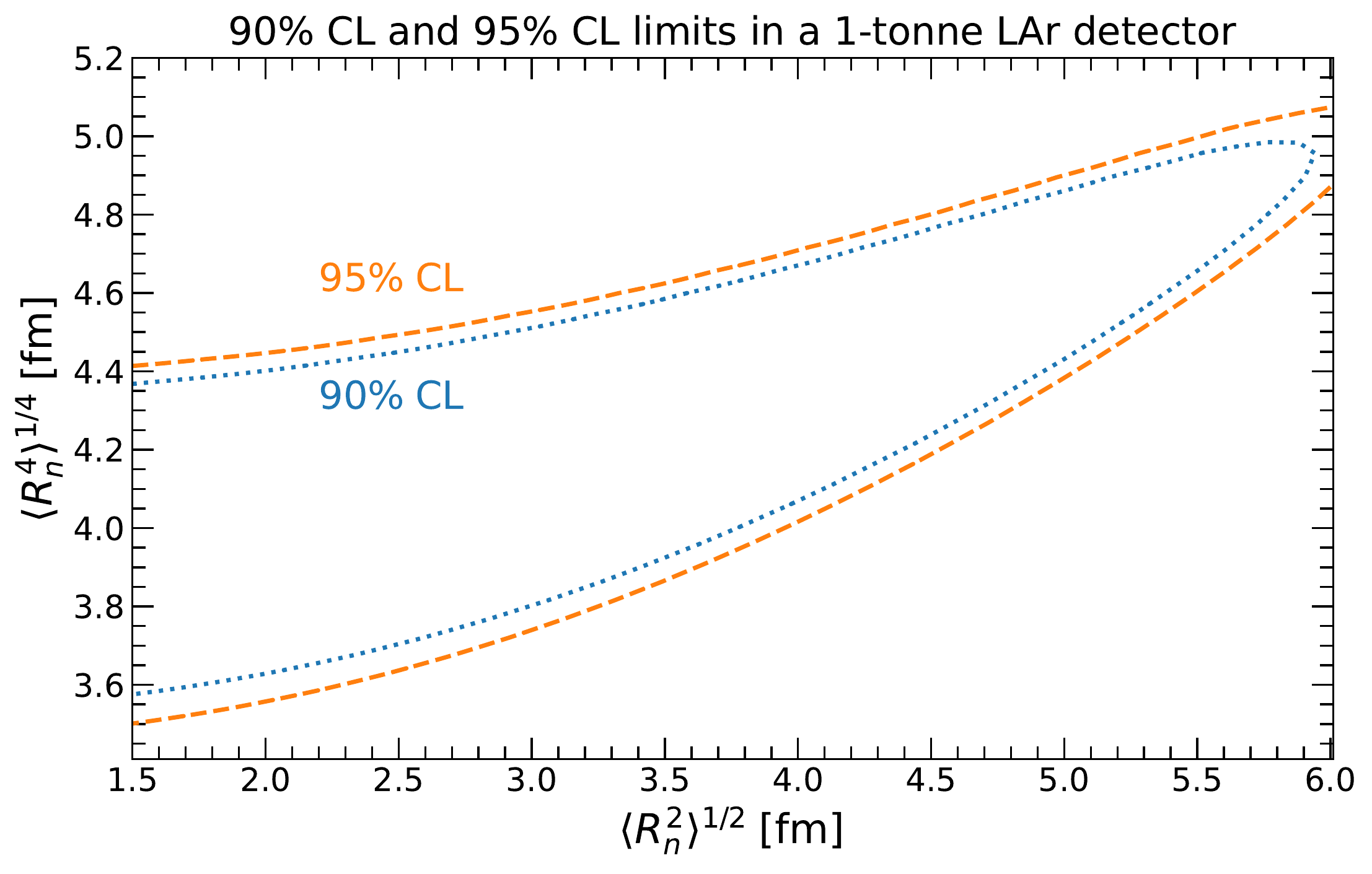}
  \caption{\textbf{Left graph}: Toy experiment data used in the
    model-independent analysis generated by fixing
    $R_n=R_p+0.1\,\text{fm}$ and
    $\langle R_n^4\rangle=210.3\,\text{fm}^4$, the latter obtained
    from the results in Eq.~(\ref{eq:four-moment-H-KN-F}) in the Helm
    parametrization case. \textbf{Right graph}: 90\% and 95\% CL
    isocontours in the neutron distribution rms radius,
    $R_n=\sqrt{\langle R_n^2\rangle}$, and fourth moment,
    $\sqrt{\langle R_n^4\rangle}$, plane.}
  \label{fig:model-indepdent-results}
\end{figure*}
Results for cases \textbf{(i)} and \textbf{(ii)} are shown in
Fig. \ref{fig:toy_experiments_signals}. Top graphs display results for
the toy experiments, while those at the bottom the results of the
$\chi^2$ analyses. Case \textbf{(i)} leads to a fit for $R_\text{W}$
from which we find at the $90\%\,$CL the following result
\begin{equation}
  \label{eq:RW_fit}
  R_\text{W} = 3.522_{-0.474}^{+0.449}\,\text{fm}\ .
\end{equation}
The point-neutron distribution rms radius can then be extracted from
this result with the aid of
Eq. (\ref{eq:EW-charge-radius-neutron-skin}). To do so one should---in
principle---take into account uncertainties on the single-nucleon
electromagnetic mean-square radii as well as on the $^{40}$Ar nuclear
charge radius. These uncertainties are given by
$\Delta \sqrt{r_p^2}=4.0\times 10^{-4}\,$fm,
$\Delta r_n^2=1.7\times 10^{-3}\,\text{fm}^2$ \cite{Zyla:2020zbs} and
$\Delta R_\text{C}=1.8\times 10^{-2}\,$fm \cite{Angeli:2013epw} and so
can be safely neglected when compared with those from $R_\text{W}$ in
Eq. (\ref{eq:RW_fit}). Results for $R_n$ at the $90\%\,$CL thus read
\begin{equation}
  \label{eq:Rn_from_RW}
  R_n = 3.428^{+0.434}_{-0.458}\,\text{fm}\ .
\end{equation}
In case \textbf{(ii)}, \textit{c'est-\`a-dire} using
Eq. (\ref{eq:EW-FF-approx}) for the weak-charge form factor, allows
fitting $R_n$ directly. In this case the diffraction radius in the
proton and neutron elastic vector form factors are fixed from $R_p$
and $R_n$. The former fixed from the nuclear-charge radius and
Eq. (\ref{eq:EM-EW-charge-radii}) (using central values for the
relevant quantities), while the latter varying over the range
$[2.6,4.1]\,\text{fm}$. At the $90\%\,$CL we find
\begin{equation}
  \label{eq:Rn_from_FW_approx}
  R_n = 3.457^{+0.492}_{-0.611}\,\text{fm}\ .
\end{equation}
Since the systematic uncertainty as well as the BRN and SS backgrounds
for both analyses have been equally fixed and the toy experiments
signals have been generated under the assumptions that define each
case, differences in the results in Eqs. (\ref{eq:Rn_from_RW}) and
(\ref{eq:Rn_from_FW_approx}) can be only attributed to theoretical
assumptions. Parametrizing the weak-charge form factor \`a la Helm,
$R_n$ can be determined with a $\sim 14\%$ precision. Using
Eq. (\ref{eq:EW-FF-approx})---including nucleon form factor terms up
to order $Q^2$---allows a determination at the $18\%$ level. We then
conclude that both procedures seem to produce results with comparable
levels of precision, though with a slight improvement if a
parametrization of the weak-charge form factor is used \footnote{Note
  that in the procedure based on expansion (\ref{eq:EW-FF-approx}) the
  elastic vector proton and neutron form factors have been as well
  parametrized using the Helm prescription. Thus this analysis is not
  a comparison among form factor parametrizations, but rather a
  comparison between results derived using a parametrization for
  $F_\text{W}$ (Helm, for definitiveness) and an expansion for
  $F_\text{W}$. Implications of different parametrizations (Helm,
  Klein-Nystrand and symmetrized Fermi function) on the CE$\nu$NS
  event rate have been analyzed in
  Ref. \cite{AristizabalSierra:2019zmy}}. In
Sec.~\ref{sec:quantitative-differences} we found that numerical
deviations from one procedure or the other could be as large as
$\sim 5\%$ in the relevant transferred momentum range. We attribute
the differences found here in the extraction of $R_n$ to that fact.

There is of course the question of whether precision can be improved
by improving upon the BRN and SS backgrounds. To assess that question
we have run to extra analyses, assuming $B_j=1.0\times N_j^\text{Exp}$
and $B_j=0.1\times N_j^\text{Exp}$. The results can be seen in both
graphs in Fig.~\ref{fig:toy_experiments_signals}. In the case in which
the Helm parametrization is used for the weak-charge form factor,
precision in the extraction of $R_n$ improves at the $5\%$ and $1\%$
levels, respectively. In the case in which $F_\text{W}$ is decomposed
in terms of elastic vector proton and neutron form factors, instead,
at the $8\%$ and $5\%$ levels.

One might wonder whether this conclusion holds as well for heavy
nuclei. Although we have not done an analysis for such a case
(e.g. for cesium), we find no reason why this should not be the
case. Of course nuclear parameters will change, but given the
discussion in the previous Sections we expect the trend to be valid in
this case too.

We now turn to case \textbf{(iii)}, for which for the calculation we
do not include nucleon form factors corrections in
Eq.~(\ref{eq:EW-FF-approx}) and expand the elastic vector proton and
neutron form factors in terms of even moments up to order $Q^4$ (as
required from the discussion in
Sec. \ref{sec:model-ind-vs-model-dep}). Such procedure leads to
\begin{align}
  \label{eq:FW_moments_expansion}
  F_W&=\frac{1}{Q_W}\left[
       Zg_V^p\left(1 - \frac{Q^2}{3!}R_p^2 + \frac{Q^4}{5!} 
       \langle R_p^4\rangle\right)\right .
       \nonumber\\
     &+ \left . Ng_V^n\left(1 - \frac{Q^2}{3!} R_n^2 + \frac{Q^4}{5!} 
       \langle R_n^4\rangle\right) \right]\ .
\end{align}
This expression depends---in principle---upon three free parameters:
The point-neutron distribution mean-square radius and the proton and
neutron fourth moments. However, the contribution controlled by
$\langle R^4_p\rangle$ can be safely neglected. This can be readily
checked by sticking---for this matter---to the Helm parametrization,
using then the corresponding expression in
Eq. (\ref{eq:four-moment-H-KN-F}) and evaluating the proton
$Q^4$-to-$Q^2$ ratio. Doing so one finds
$2\times 10^{-5}(Q/\text{MeV})^2$, which combined with the fact that
$Q\lesssim 50\,\text{MeV}$ and that the cross section is dominated by
the neutron contribution reduces the calculation to a two-parameter
problem.

Dropping the proton fourth moment one can then determine from data the
neutron distribution mean-square radius and its fourth moment. To
generate the toy experiment data we fix $\langle R_n^4\rangle$ with
the aid of Eq. (\ref{eq:four-moment-H-KN-F}) assuming the Helm
parametrization. Results for that pseudo-data are displayed in the
left graph in Fig.~\ref{fig:model-indepdent-results}. We then proceed
with the least-square analysis by varying
$R_n\equiv \sqrt{\langle R_n^2\rangle}$ and $\langle R_n^4\rangle$.
Results of this calculation in the
$\sqrt{\langle R_n^2\rangle}$-$\langle R_n^4\rangle^{1/4}$ plane are
shown in the right graph in Fig. \ref{fig:model-indepdent-results}.
One can see that under the same background and detector assumptions
that those used in the previous two cases, only an upper limit on
$R_n$ can be derived. At the $90\%$CL one gets
\begin{equation}
  \label{eq:R_n_model_independent}
  R_n\lesssim 5.9\,\text{fm}\ .
\end{equation}
The result for the fourth moment is instead constrained to an interval
(at both the $90\%$ and $95\%$ CLs), although wide. Fitting $R_n$
through the model-independent approach has---of course---the advantage
of not relying on particular nuclear physics assumptions, but implies
fitting two parameters rather than one. That is why, with the
statistics and background assumptions we have adopted, only an upper
limit on $R_n$ can be placed.

In some respect this result differs from the one reported in
Ref. \cite{Patton:2012jr}, in particular in the shape of the available
$90\%$ and $95\%$ CLs contours. It seems to us these differences are
expected because of the following reasons: (a) Our analysis includes
the proton contribution (which might become important in regions of
parameter space where
$R_n/\langle R_n^4\rangle^{1/2}\simeq Q/\sqrt{20}$), (b) our analysis
involves a 3.5 less statistics (a 1-tonne rather than 3.5-tonne LAr
detector), (c) our systematic and statistical uncertainties are much
larger (in particular the statistical uncertainty which includes a
background hypothesis that exceeds the signal by a factor 10), (d) the
statistical methods employed in the extraction of the relevant
quantities.

Results from the three different approaches we have employed here thus
seem to favor the inclusion of nucleon form factors $Q$-dependent
terms and the decomposition of the weak-charge form factor in terms of
elastic vector proton and neutron form factors, as given in
Eq.~(\ref{eq:EW-FF-approx}). With such an approach a percent
determination of the neutron distribution rms radius seems achievable,
as shown in Eq.~(\ref{eq:Rn_from_FW_approx}). Improving the
statistical uncertainty by improving upon BRN and SS backgrounds may
lead to $\sim 1\%$ determination of $R_n$.
\section{Conclusions}
\label{sec:conclusions}
We have quantified uncertainties on the extraction of point-neutron
distributions mean-square radii from CE$\nu$NS data. Our analysis is
motivated by future $\mathcal{O}(\text{tonne})$-scale CE$\nu$NS
detectors which will deliver thousands of events per year with
well-controlled systematic and statistical
uncertainties. Theoretically, uncertainties are encoded in the
weak-charge form factor, for which a variety of effects and
parametrization approaches have different impact.

Here we have quantified at the weak-charge form factor level
uncertainties due to: (i) The absence/presence of single-nucleon
electromagnetic mean-square radii in the determination of the
point-proton distribution mean-square radius, (ii) $Q$-dependent
nucleon form factor terms up to order $Q^2$, (iii) parametrizations of
the elastic vector proton and neutron form factors, (iv)
parametrizations of the weak-charge form factor, (v) model-independent
series expansions of the elastic vector proton and neutron form
factors.

At the weak-charge level we have found that the inclusion of
single-nucleon electromagnetic mean-square radii in the determination
of the point-proton distribution mean-square radius has a per mille
impact. Thus showing that taking the point-proton distribution
mean-square radius equal to the nuclear charge radius is precise
enough. Inclusion of nucleon form factors $Q$-dependent terms,
instead, has a percent effect. Comparison of results from
decompositions of the weak-charge form factor in terms of elastic
vector proton and neutron form factors with weak-charge form factor
parametrizations shows a wider uncertainty, that can rise up to the
order of $\sim 5\%$ ($\sim 10\%$) in the relevant transferred momentum
range for light (heavy) nuclei. Comparison of results from
decompositions of the weak-charge form factor in terms of elastic
vector proton and neutron form factors with series expansions of the
elastic vector proton and neutron form factors shows uncertainties of
the same order.

To better understand the impact of these effects we have considered
pseudo-data from a one-tonne LAr detector with substantial BRN and SS
backgrounds, though suppressed compared with those in current LAr
data. Our results show that weak-charge form factor parametrizations
and decompositions of the weak-charge form factor in terms of elastic
vector proton and neutron form factors lead to the same level of
accuracy. An indication that under the experimental conditions the
analysis has been carried out, both approaches lead to
indistinguishable results. Suppressed BRN and SS backgrounds seem to
enable reaching $\sim 1\%$ accuracies in the extraction of $R_n$.
\section*{Acknowledgments}
The author warmly thank Dimitrios Papoulias and Jorge Piekarewicz for
very useful comments on the manuscript.  He thank as well the
``Universidad de Antioquia'' Physics Department for its hospitality
during the completion of this work and ANID for financial support
through grant ``Fondecyt Regular'' 1221445.
\bibliography{references}

\begin{thebibliography}{57}
\expandafter\ifx\csname natexlab\endcsname\relax\def\natexlab#1{#1}\fi
\expandafter\ifx\csname bibnamefont\endcsname\relax
  \def\bibnamefont#1{#1}\fi
\expandafter\ifx\csname bibfnamefont\endcsname\relax
  \def\bibfnamefont#1{#1}\fi
\expandafter\ifx\csname citenamefont\endcsname\relax
  \def\citenamefont#1{#1}\fi
\expandafter\ifx\csname url\endcsname\relax
  \def\url#1{\texttt{#1}}\fi
\expandafter\ifx\csname urlprefix\endcsname\relax\def\urlprefix{URL }\fi
\providecommand{\bibinfo}[2]{#2}
\providecommand{\eprint}[2][]{\url{#2}}

\bibitem[{\citenamefont{Akimov et~al.}(2017)}]{Akimov:2017ade}
\bibinfo{author}{\bibfnamefont{D.}~\bibnamefont{Akimov}} \bibnamefont{et~al.}
  (\bibinfo{collaboration}{COHERENT}), \bibinfo{journal}{Science}
  (\bibinfo{year}{2017}), \eprint{1708.01294}.

\bibitem[{\citenamefont{Akimov et~al.}(2022)}]{COHERENT:2021xmm}
\bibinfo{author}{\bibfnamefont{D.}~\bibnamefont{Akimov}} \bibnamefont{et~al.}
  (\bibinfo{collaboration}{COHERENT}), \bibinfo{journal}{Phys. Rev. Lett.}
  \textbf{\bibinfo{volume}{129}}, \bibinfo{pages}{081801}
  (\bibinfo{year}{2022}), \eprint{2110.07730}.

\bibitem[{\citenamefont{Akimov et~al.}(2020{\natexlab{a}})}]{Akimov:2020czh}
\bibinfo{author}{\bibfnamefont{D.}~\bibnamefont{Akimov}} \bibnamefont{et~al.}
  (\bibinfo{collaboration}{COHERENT}) (\bibinfo{year}{2020}{\natexlab{a}}),
  \eprint{2006.12659}.

\bibitem[{\citenamefont{Cadeddu
  et~al.}(2018{\natexlab{a}})\citenamefont{Cadeddu, Giunti, Li, and
  Zhang}}]{Cadeddu:2017etk}
\bibinfo{author}{\bibfnamefont{M.}~\bibnamefont{Cadeddu}},
  \bibinfo{author}{\bibfnamefont{C.}~\bibnamefont{Giunti}},
  \bibinfo{author}{\bibfnamefont{Y.~F.} \bibnamefont{Li}}, \bibnamefont{and}
  \bibinfo{author}{\bibfnamefont{Y.~Y.} \bibnamefont{Zhang}},
  \bibinfo{journal}{Phys. Rev. Lett.} \textbf{\bibinfo{volume}{120}},
  \bibinfo{pages}{072501} (\bibinfo{year}{2018}{\natexlab{a}}),
  \eprint{1710.02730}.

\bibitem[{\citenamefont{Papoulias}(2020)}]{Papoulias:2019txv}
\bibinfo{author}{\bibfnamefont{D.~K.} \bibnamefont{Papoulias}},
  \bibinfo{journal}{Phys. Rev. D} \textbf{\bibinfo{volume}{102}},
  \bibinfo{pages}{113004} (\bibinfo{year}{2020}), \eprint{1907.11644}.

\bibitem[{\citenamefont{Miranda et~al.}(2020)\citenamefont{Miranda, Papoulias,
  Sanchez~Garcia, Sanders, T\'ortola, and Valle}}]{Miranda:2020tif}
\bibinfo{author}{\bibfnamefont{O.}~\bibnamefont{Miranda}},
  \bibinfo{author}{\bibfnamefont{D.}~\bibnamefont{Papoulias}},
  \bibinfo{author}{\bibfnamefont{G.}~\bibnamefont{Sanchez~Garcia}},
  \bibinfo{author}{\bibfnamefont{O.}~\bibnamefont{Sanders}},
  \bibinfo{author}{\bibfnamefont{M.}~\bibnamefont{T\'ortola}},
  \bibnamefont{and} \bibinfo{author}{\bibfnamefont{J.}~\bibnamefont{Valle}},
  \bibinfo{journal}{JHEP} \textbf{\bibinfo{volume}{05}}, \bibinfo{pages}{130}
  (\bibinfo{year}{2020}), \eprint{2003.12050}.

\bibitem[{\citenamefont{Cadeddu
  et~al.}(2021{\natexlab{a}})\citenamefont{Cadeddu, Cargioli, Dordei, Giunti,
  Li, Picciau, Ternes, and Zhang}}]{Cadeddu:2021ijh}
\bibinfo{author}{\bibfnamefont{M.}~\bibnamefont{Cadeddu}},
  \bibinfo{author}{\bibfnamefont{N.}~\bibnamefont{Cargioli}},
  \bibinfo{author}{\bibfnamefont{F.}~\bibnamefont{Dordei}},
  \bibinfo{author}{\bibfnamefont{C.}~\bibnamefont{Giunti}},
  \bibinfo{author}{\bibfnamefont{Y.~F.} \bibnamefont{Li}},
  \bibinfo{author}{\bibfnamefont{E.}~\bibnamefont{Picciau}},
  \bibinfo{author}{\bibfnamefont{C.~A.} \bibnamefont{Ternes}},
  \bibnamefont{and} \bibinfo{author}{\bibfnamefont{Y.~Y.} \bibnamefont{Zhang}},
  \bibinfo{journal}{Phys. Rev. C} \textbf{\bibinfo{volume}{104}},
  \bibinfo{pages}{065502} (\bibinfo{year}{2021}{\natexlab{a}}),
  \eprint{2102.06153}.

\bibitem[{\citenamefont{Coloma et~al.}(2020)\citenamefont{Coloma, Esteban,
  Gonzalez-Garcia, and Menendez}}]{Coloma:2020nhf}
\bibinfo{author}{\bibfnamefont{P.}~\bibnamefont{Coloma}},
  \bibinfo{author}{\bibfnamefont{I.}~\bibnamefont{Esteban}},
  \bibinfo{author}{\bibfnamefont{M.~C.} \bibnamefont{Gonzalez-Garcia}},
  \bibnamefont{and} \bibinfo{author}{\bibfnamefont{J.}~\bibnamefont{Menendez}},
  \bibinfo{journal}{JHEP} \textbf{\bibinfo{volume}{08}}, \bibinfo{pages}{030}
  (\bibinfo{year}{2020}), \eprint{2006.08624}.

\bibitem[{\citenamefont{De~Romeri et~al.}(2022)\citenamefont{De~Romeri,
  Miranda, Papoulias, Sanchez~Garcia, T\'ortola, and Valle}}]{DeRomeri:2022twg}
\bibinfo{author}{\bibfnamefont{V.}~\bibnamefont{De~Romeri}},
  \bibinfo{author}{\bibfnamefont{O.~G.} \bibnamefont{Miranda}},
  \bibinfo{author}{\bibfnamefont{D.~K.} \bibnamefont{Papoulias}},
  \bibinfo{author}{\bibfnamefont{G.}~\bibnamefont{Sanchez~Garcia}},
  \bibinfo{author}{\bibfnamefont{M.}~\bibnamefont{T\'ortola}},
  \bibnamefont{and} \bibinfo{author}{\bibfnamefont{J.~W.~F.}
  \bibnamefont{Valle}} (\bibinfo{year}{2022}), \eprint{2211.11905}.

\bibitem[{\citenamefont{Papoulias and
  Kosmas}(2018{\natexlab{a}})}]{Papoulias:2017qdn}
\bibinfo{author}{\bibfnamefont{D.~K.} \bibnamefont{Papoulias}}
  \bibnamefont{and} \bibinfo{author}{\bibfnamefont{T.~S.}
  \bibnamefont{Kosmas}}, \bibinfo{journal}{Phys. Rev. D}
  \textbf{\bibinfo{volume}{97}}, \bibinfo{pages}{033003}
  (\bibinfo{year}{2018}{\natexlab{a}}), \eprint{1711.09773}.

\bibitem[{\citenamefont{Liao and Marfatia}(2017)}]{Liao:2017uzy}
\bibinfo{author}{\bibfnamefont{J.}~\bibnamefont{Liao}} \bibnamefont{and}
  \bibinfo{author}{\bibfnamefont{D.}~\bibnamefont{Marfatia}},
  \bibinfo{journal}{Phys. Lett.} \textbf{\bibinfo{volume}{B775}},
  \bibinfo{pages}{54} (\bibinfo{year}{2017}), \eprint{1708.04255}.

\bibitem[{\citenamefont{Aristizabal~Sierra
  et~al.}(2018)\citenamefont{Aristizabal~Sierra, De~Romeri, and
  Rojas}}]{AristizabalSierra:2018eqm}
\bibinfo{author}{\bibfnamefont{D.}~\bibnamefont{Aristizabal~Sierra}},
  \bibinfo{author}{\bibfnamefont{V.}~\bibnamefont{De~Romeri}},
  \bibnamefont{and} \bibinfo{author}{\bibfnamefont{N.}~\bibnamefont{Rojas}},
  \bibinfo{journal}{Phys. Rev.} \textbf{\bibinfo{volume}{D98}},
  \bibinfo{pages}{075018} (\bibinfo{year}{2018}), \eprint{1806.07424}.

\bibitem[{\citenamefont{Aristizabal~Sierra
  et~al.}(2019{\natexlab{a}})\citenamefont{Aristizabal~Sierra, De~Romeri, and
  Rojas}}]{AristizabalSierra:2019ufd}
\bibinfo{author}{\bibfnamefont{D.}~\bibnamefont{Aristizabal~Sierra}},
  \bibinfo{author}{\bibfnamefont{V.}~\bibnamefont{De~Romeri}},
  \bibnamefont{and} \bibinfo{author}{\bibfnamefont{N.}~\bibnamefont{Rojas}},
  \bibinfo{journal}{JHEP} \textbf{\bibinfo{volume}{09}}, \bibinfo{pages}{069}
  (\bibinfo{year}{2019}{\natexlab{a}}), \eprint{1906.01156}.

\bibitem[{\citenamefont{Aristizabal~Sierra
  et~al.}(2019{\natexlab{b}})\citenamefont{Aristizabal~Sierra, Dutta, Liao, and
  Strigari}}]{AristizabalSierra:2019ykk}
\bibinfo{author}{\bibfnamefont{D.}~\bibnamefont{Aristizabal~Sierra}},
  \bibinfo{author}{\bibfnamefont{B.}~\bibnamefont{Dutta}},
  \bibinfo{author}{\bibfnamefont{S.}~\bibnamefont{Liao}}, \bibnamefont{and}
  \bibinfo{author}{\bibfnamefont{L.~E.} \bibnamefont{Strigari}},
  \bibinfo{journal}{JHEP} \textbf{\bibinfo{volume}{12}}, \bibinfo{pages}{124}
  (\bibinfo{year}{2019}{\natexlab{b}}), \eprint{1910.12437}.

\bibitem[{\citenamefont{Dutta et~al.}(2019)\citenamefont{Dutta, Liao, Sinha,
  and Strigari}}]{Dutta:2019eml}
\bibinfo{author}{\bibfnamefont{B.}~\bibnamefont{Dutta}},
  \bibinfo{author}{\bibfnamefont{S.}~\bibnamefont{Liao}},
  \bibinfo{author}{\bibfnamefont{S.}~\bibnamefont{Sinha}}, \bibnamefont{and}
  \bibinfo{author}{\bibfnamefont{L.~E.} \bibnamefont{Strigari}},
  \bibinfo{journal}{Phys. Rev. Lett.} \textbf{\bibinfo{volume}{123}},
  \bibinfo{pages}{061801} (\bibinfo{year}{2019}), \eprint{1903.10666}.

\bibitem[{\citenamefont{Cadeddu
  et~al.}(2021{\natexlab{b}})\citenamefont{Cadeddu, Cargioli, Dordei, Giunti,
  Li, Picciau, and Zhang}}]{Cadeddu:2020nbr}
\bibinfo{author}{\bibfnamefont{M.}~\bibnamefont{Cadeddu}},
  \bibinfo{author}{\bibfnamefont{N.}~\bibnamefont{Cargioli}},
  \bibinfo{author}{\bibfnamefont{F.}~\bibnamefont{Dordei}},
  \bibinfo{author}{\bibfnamefont{C.}~\bibnamefont{Giunti}},
  \bibinfo{author}{\bibfnamefont{Y.~F.} \bibnamefont{Li}},
  \bibinfo{author}{\bibfnamefont{E.}~\bibnamefont{Picciau}}, \bibnamefont{and}
  \bibinfo{author}{\bibfnamefont{Y.~Y.} \bibnamefont{Zhang}},
  \bibinfo{journal}{JHEP} \textbf{\bibinfo{volume}{01}}, \bibinfo{pages}{116}
  (\bibinfo{year}{2021}{\natexlab{b}}), \eprint{2008.05022}.

\bibitem[{\citenamefont{Banerjee et~al.}(2021)\citenamefont{Banerjee, Dutta,
  and Roy}}]{Banerjee:2021laz}
\bibinfo{author}{\bibfnamefont{H.}~\bibnamefont{Banerjee}},
  \bibinfo{author}{\bibfnamefont{B.}~\bibnamefont{Dutta}}, \bibnamefont{and}
  \bibinfo{author}{\bibfnamefont{S.}~\bibnamefont{Roy}},
  \bibinfo{journal}{Phys. Rev. D} \textbf{\bibinfo{volume}{104}},
  \bibinfo{pages}{015015} (\bibinfo{year}{2021}), \eprint{2103.10196}.

\bibitem[{\citenamefont{Abdullah et~al.}(2018)\citenamefont{Abdullah, Dent,
  Dutta, Kane, Liao, and Strigari}}]{Abdullah:2018ykz}
\bibinfo{author}{\bibfnamefont{M.}~\bibnamefont{Abdullah}},
  \bibinfo{author}{\bibfnamefont{J.~B.} \bibnamefont{Dent}},
  \bibinfo{author}{\bibfnamefont{B.}~\bibnamefont{Dutta}},
  \bibinfo{author}{\bibfnamefont{G.~L.} \bibnamefont{Kane}},
  \bibinfo{author}{\bibfnamefont{S.}~\bibnamefont{Liao}}, \bibnamefont{and}
  \bibinfo{author}{\bibfnamefont{L.~E.} \bibnamefont{Strigari}},
  \bibinfo{journal}{Phys. Rev. D} \textbf{\bibinfo{volume}{98}},
  \bibinfo{pages}{015005} (\bibinfo{year}{2018}), \eprint{1803.01224}.

\bibitem[{\citenamefont{Shoemaker}(2017)}]{Shoemaker:2017lzs}
\bibinfo{author}{\bibfnamefont{I.~M.} \bibnamefont{Shoemaker}},
  \bibinfo{journal}{Phys. Rev.} \textbf{\bibinfo{volume}{D95}},
  \bibinfo{pages}{115028} (\bibinfo{year}{2017}), \eprint{1703.05774}.

\bibitem[{\citenamefont{Ge and Shoemaker}(2017)}]{Ge:2017mcq}
\bibinfo{author}{\bibfnamefont{S.-F.} \bibnamefont{Ge}} \bibnamefont{and}
  \bibinfo{author}{\bibfnamefont{I.~M.} \bibnamefont{Shoemaker}}
  (\bibinfo{year}{2017}), \eprint{1710.10889}.

\bibitem[{\citenamefont{Denton et~al.}(2018)\citenamefont{Denton, Farzan, and
  Shoemaker}}]{Denton:2018xmq}
\bibinfo{author}{\bibfnamefont{P.~B.} \bibnamefont{Denton}},
  \bibinfo{author}{\bibfnamefont{Y.}~\bibnamefont{Farzan}}, \bibnamefont{and}
  \bibinfo{author}{\bibfnamefont{I.~M.} \bibnamefont{Shoemaker}},
  \bibinfo{journal}{JHEP} \textbf{\bibinfo{volume}{07}}, \bibinfo{pages}{037}
  (\bibinfo{year}{2018}), \eprint{1804.03660}.

\bibitem[{\citenamefont{Denton and Gehrlein}(2021)}]{Denton:2020hop}
\bibinfo{author}{\bibfnamefont{P.~B.} \bibnamefont{Denton}} \bibnamefont{and}
  \bibinfo{author}{\bibfnamefont{J.}~\bibnamefont{Gehrlein}},
  \bibinfo{journal}{JHEP} \textbf{\bibinfo{volume}{04}}, \bibinfo{pages}{266}
  (\bibinfo{year}{2021}), \eprint{2008.06062}.

\bibitem[{\citenamefont{Coloma et~al.}(2017)\citenamefont{Coloma,
  Gonzalez-Garcia, Maltoni, and Schwetz}}]{Coloma:2017ncl}
\bibinfo{author}{\bibfnamefont{P.}~\bibnamefont{Coloma}},
  \bibinfo{author}{\bibfnamefont{M.~C.} \bibnamefont{Gonzalez-Garcia}},
  \bibinfo{author}{\bibfnamefont{M.}~\bibnamefont{Maltoni}}, \bibnamefont{and}
  \bibinfo{author}{\bibfnamefont{T.}~\bibnamefont{Schwetz}}
  (\bibinfo{year}{2017}), \eprint{1708.02899}.

\bibitem[{\citenamefont{Coloma et~al.}(2019)\citenamefont{Coloma, Esteban,
  Gonzalez-Garcia, and Maltoni}}]{Coloma:2019mbs}
\bibinfo{author}{\bibfnamefont{P.}~\bibnamefont{Coloma}},
  \bibinfo{author}{\bibfnamefont{I.}~\bibnamefont{Esteban}},
  \bibinfo{author}{\bibfnamefont{M.~C.} \bibnamefont{Gonzalez-Garcia}},
  \bibnamefont{and} \bibinfo{author}{\bibfnamefont{M.}~\bibnamefont{Maltoni}}
  (\bibinfo{year}{2019}), \eprint{1911.09109}.

\bibitem[{\citenamefont{Cadeddu
  et~al.}(2018{\natexlab{b}})\citenamefont{Cadeddu, Giunti, Kouzakov, Li,
  Zhang, and Studenikin}}]{Cadeddu:2018dux}
\bibinfo{author}{\bibfnamefont{M.}~\bibnamefont{Cadeddu}},
  \bibinfo{author}{\bibfnamefont{C.}~\bibnamefont{Giunti}},
  \bibinfo{author}{\bibfnamefont{K.~A.} \bibnamefont{Kouzakov}},
  \bibinfo{author}{\bibfnamefont{Y.-F.} \bibnamefont{Li}},
  \bibinfo{author}{\bibfnamefont{Y.-Y.} \bibnamefont{Zhang}}, \bibnamefont{and}
  \bibinfo{author}{\bibfnamefont{A.~I.} \bibnamefont{Studenikin}},
  \bibinfo{journal}{Phys. Rev. D} \textbf{\bibinfo{volume}{98}},
  \bibinfo{pages}{113010} (\bibinfo{year}{2018}{\natexlab{b}}),
  \bibinfo{note}{[Erratum: Phys.Rev.D 101, 059902 (2020)]},
  \eprint{1810.05606}.

\bibitem[{\citenamefont{Miranda et~al.}(2019)\citenamefont{Miranda, Papoulias,
  T\'ortola, and Valle}}]{Miranda:2019wdy}
\bibinfo{author}{\bibfnamefont{O.~G.} \bibnamefont{Miranda}},
  \bibinfo{author}{\bibfnamefont{D.~K.} \bibnamefont{Papoulias}},
  \bibinfo{author}{\bibfnamefont{M.}~\bibnamefont{T\'ortola}},
  \bibnamefont{and} \bibinfo{author}{\bibfnamefont{J.~W.~F.}
  \bibnamefont{Valle}}, \bibinfo{journal}{JHEP} \textbf{\bibinfo{volume}{07}},
  \bibinfo{pages}{103} (\bibinfo{year}{2019}), \eprint{1905.03750}.

\bibitem[{\citenamefont{Khan}(2023)}]{Khan:2022akj}
\bibinfo{author}{\bibfnamefont{A.~N.} \bibnamefont{Khan}},
  \bibinfo{journal}{Nucl. Phys. B} \textbf{\bibinfo{volume}{986}},
  \bibinfo{pages}{116064} (\bibinfo{year}{2023}), \eprint{2201.10578}.

\bibitem[{\citenamefont{{D. Anderson et al.}}(2015)}]{ref:STS}
\bibinfo{author}{\bibnamefont{{D. Anderson et al.}}},
  \emph{\bibinfo{title}{{Technical Design Report Second Target Station}}},
  \bibinfo{howpublished}{\url{https://neutrons.ornl.gov/sites/default/files/SNS_STS_Technical_Design_Report_2015-01.pdf}}
  (\bibinfo{year}{2015}).

\bibitem[{\citenamefont{Asaadi et~al.}(2022)}]{Asaadi:2022ojm}
\bibinfo{author}{\bibfnamefont{J.}~\bibnamefont{Asaadi}} \bibnamefont{et~al.},
  in \emph{\bibinfo{booktitle}{{2022 Snowmass Summer Study}}}
  (\bibinfo{year}{2022}), \eprint{2209.02883}.

\bibitem[{\citenamefont{Akimov et~al.}(2020{\natexlab{b}})}]{COHERENT:2019kwz}
\bibinfo{author}{\bibfnamefont{D.}~\bibnamefont{Akimov}} \bibnamefont{et~al.}
  (\bibinfo{collaboration}{COHERENT}), \bibinfo{journal}{Phys. Rev. D}
  \textbf{\bibinfo{volume}{102}}, \bibinfo{pages}{052007}
  (\bibinfo{year}{2020}{\natexlab{b}}), \eprint{1911.06422}.

\bibitem[{cap(2018)}]{captain-mills}
\emph{\bibinfo{title}{{C}oherent {C}aptain-{M}ills ({CCM}) {E}xperiment}},
  \bibinfo{howpublished}{\url{https://p25ext.lanl.gov/~lee/CaptainMills/}}
  (\bibinfo{year}{2018}).

\bibitem[{\citenamefont{Aguilar-Arevalo
  et~al.}(2022{\natexlab{a}})}]{CCM:2021leg}
\bibinfo{author}{\bibfnamefont{A.~A.} \bibnamefont{Aguilar-Arevalo}}
  \bibnamefont{et~al.} (\bibinfo{collaboration}{CCM}), \bibinfo{journal}{Phys.
  Rev. D} \textbf{\bibinfo{volume}{106}}, \bibinfo{pages}{012001}
  (\bibinfo{year}{2022}{\natexlab{a}}), \eprint{2105.14020}.

\bibitem[{\citenamefont{Aguilar-Arevalo
  et~al.}(2022{\natexlab{b}})}]{CCM:2021yzc}
\bibinfo{author}{\bibfnamefont{A.~A.} \bibnamefont{Aguilar-Arevalo}}
  \bibnamefont{et~al.} (\bibinfo{collaboration}{CCM}), \bibinfo{journal}{Phys.
  Rev. Lett.} \textbf{\bibinfo{volume}{129}}, \bibinfo{pages}{021801}
  (\bibinfo{year}{2022}{\natexlab{b}}), \eprint{2109.14146}.

\bibitem[{\citenamefont{Tomalak et~al.}(2021)\citenamefont{Tomalak, Machado,
  Pandey, and Plestid}}]{Tomalak:2020zfh}
\bibinfo{author}{\bibfnamefont{O.}~\bibnamefont{Tomalak}},
  \bibinfo{author}{\bibfnamefont{P.}~\bibnamefont{Machado}},
  \bibinfo{author}{\bibfnamefont{V.}~\bibnamefont{Pandey}}, \bibnamefont{and}
  \bibinfo{author}{\bibfnamefont{R.}~\bibnamefont{Plestid}},
  \bibinfo{journal}{JHEP} \textbf{\bibinfo{volume}{02}}, \bibinfo{pages}{097}
  (\bibinfo{year}{2021}), \eprint{2011.05960}.

\bibitem[{\citenamefont{Papoulias and
  Kosmas}(2018{\natexlab{b}})}]{Kosmas:2017tsq}
\bibinfo{author}{\bibfnamefont{D.~K.} \bibnamefont{Papoulias}}
  \bibnamefont{and} \bibinfo{author}{\bibfnamefont{T.~S.}
  \bibnamefont{Kosmas}}, \bibinfo{journal}{Phys. Rev.}
  \textbf{\bibinfo{volume}{D97}}, \bibinfo{pages}{033003}
  (\bibinfo{year}{2018}{\natexlab{b}}), \eprint{1711.09773}.

\bibitem[{\citenamefont{Aristizabal~Sierra
  et~al.}(2019{\natexlab{c}})\citenamefont{Aristizabal~Sierra, Liao, and
  Marfatia}}]{AristizabalSierra:2019zmy}
\bibinfo{author}{\bibfnamefont{D.}~\bibnamefont{Aristizabal~Sierra}},
  \bibinfo{author}{\bibfnamefont{J.}~\bibnamefont{Liao}}, \bibnamefont{and}
  \bibinfo{author}{\bibfnamefont{D.}~\bibnamefont{Marfatia}},
  \bibinfo{journal}{JHEP} \textbf{\bibinfo{volume}{06}}, \bibinfo{pages}{141}
  (\bibinfo{year}{2019}{\natexlab{c}}), \eprint{1902.07398}.

\bibitem[{\citenamefont{Aristizabal~Sierra
  et~al.}(2021)\citenamefont{Aristizabal~Sierra, Dutta, Kim, Snowden-Ifft, and
  Strigari}}]{AristizabalSierra:2021uob}
\bibinfo{author}{\bibfnamefont{D.}~\bibnamefont{Aristizabal~Sierra}},
  \bibinfo{author}{\bibfnamefont{B.}~\bibnamefont{Dutta}},
  \bibinfo{author}{\bibfnamefont{D.}~\bibnamefont{Kim}},
  \bibinfo{author}{\bibfnamefont{D.}~\bibnamefont{Snowden-Ifft}},
  \bibnamefont{and} \bibinfo{author}{\bibfnamefont{L.~E.}
  \bibnamefont{Strigari}}, \bibinfo{journal}{Phys. Rev. D}
  \textbf{\bibinfo{volume}{104}}, \bibinfo{pages}{033004}
  (\bibinfo{year}{2021}), \eprint{2103.10857}.

\bibitem[{\citenamefont{Patton et~al.}(2012)\citenamefont{Patton, Engel,
  McLaughlin, and Schunck}}]{Patton:2012jr}
\bibinfo{author}{\bibfnamefont{K.}~\bibnamefont{Patton}},
  \bibinfo{author}{\bibfnamefont{J.}~\bibnamefont{Engel}},
  \bibinfo{author}{\bibfnamefont{G.~C.} \bibnamefont{McLaughlin}},
  \bibnamefont{and} \bibinfo{author}{\bibfnamefont{N.}~\bibnamefont{Schunck}},
  \bibinfo{journal}{Phys. Rev. C} \textbf{\bibinfo{volume}{86}},
  \bibinfo{pages}{024612} (\bibinfo{year}{2012}), \eprint{1207.0693}.

\bibitem[{\citenamefont{Friar et~al.}(1997)\citenamefont{Friar, Martorell, and
  Sprung}}]{Friar:1997js}
\bibinfo{author}{\bibfnamefont{J.~L.} \bibnamefont{Friar}},
  \bibinfo{author}{\bibfnamefont{J.}~\bibnamefont{Martorell}},
  \bibnamefont{and} \bibinfo{author}{\bibfnamefont{D.~W.~L.}
  \bibnamefont{Sprung}}, \bibinfo{journal}{Phys. Rev. A}
  \textbf{\bibinfo{volume}{56}}, \bibinfo{pages}{4579} (\bibinfo{year}{1997}),
  \eprint{nucl-th/9707016}.

\bibitem[{\citenamefont{Horowitz and Piekarewicz}(2012)}]{Horowitz:2012we}
\bibinfo{author}{\bibfnamefont{C.~J.} \bibnamefont{Horowitz}} \bibnamefont{and}
  \bibinfo{author}{\bibfnamefont{J.}~\bibnamefont{Piekarewicz}},
  \bibinfo{journal}{Phys. Rev. C} \textbf{\bibinfo{volume}{86}},
  \bibinfo{pages}{045503} (\bibinfo{year}{2012}), \eprint{1208.2249}.

\bibitem[{\citenamefont{Ernst et~al.}(1960)\citenamefont{Ernst, Sachs, and
  Wali}}]{Ernst:1960zza}
\bibinfo{author}{\bibfnamefont{F.~J.} \bibnamefont{Ernst}},
  \bibinfo{author}{\bibfnamefont{R.~G.} \bibnamefont{Sachs}}, \bibnamefont{and}
  \bibinfo{author}{\bibfnamefont{K.~C.} \bibnamefont{Wali}},
  \bibinfo{journal}{Phys. Rev.} \textbf{\bibinfo{volume}{119}},
  \bibinfo{pages}{1105} (\bibinfo{year}{1960}).

\bibitem[{\citenamefont{Liu et~al.}(2007)\citenamefont{Liu, McKeown, and
  Ramsey-Musolf}}]{Liu:2007yi}
\bibinfo{author}{\bibfnamefont{J.}~\bibnamefont{Liu}},
  \bibinfo{author}{\bibfnamefont{R.~D.} \bibnamefont{McKeown}},
  \bibnamefont{and} \bibinfo{author}{\bibfnamefont{M.~J.}
  \bibnamefont{Ramsey-Musolf}}, \bibinfo{journal}{Phys. Rev. C}
  \textbf{\bibinfo{volume}{76}}, \bibinfo{pages}{025202}
  (\bibinfo{year}{2007}), \eprint{0706.0226}.

\bibitem[{\citenamefont{Sufian et~al.}(2017)\citenamefont{Sufian, Yang,
  Alexandru, Draper, Liang, and Liu}}]{Sufian:2016pex}
\bibinfo{author}{\bibfnamefont{R.~S.} \bibnamefont{Sufian}},
  \bibinfo{author}{\bibfnamefont{Y.-B.} \bibnamefont{Yang}},
  \bibinfo{author}{\bibfnamefont{A.}~\bibnamefont{Alexandru}},
  \bibinfo{author}{\bibfnamefont{T.}~\bibnamefont{Draper}},
  \bibinfo{author}{\bibfnamefont{J.}~\bibnamefont{Liang}}, \bibnamefont{and}
  \bibinfo{author}{\bibfnamefont{K.-F.} \bibnamefont{Liu}},
  \bibinfo{journal}{Phys. Rev. Lett.} \textbf{\bibinfo{volume}{118}},
  \bibinfo{pages}{042001} (\bibinfo{year}{2017}), \eprint{1606.07075}.

\bibitem[{\citenamefont{Helm}(1956)}]{Helm:1956zz}
\bibinfo{author}{\bibfnamefont{R.~H.} \bibnamefont{Helm}},
  \bibinfo{journal}{Phys. Rev.} \textbf{\bibinfo{volume}{104}},
  \bibinfo{pages}{1466} (\bibinfo{year}{1956}).

\bibitem[{\citenamefont{Klein and Nystrand}(1999)}]{Klein:1999qj}
\bibinfo{author}{\bibfnamefont{S.}~\bibnamefont{Klein}} \bibnamefont{and}
  \bibinfo{author}{\bibfnamefont{J.}~\bibnamefont{Nystrand}},
  \bibinfo{journal}{Phys. Rev.} \textbf{\bibinfo{volume}{C60}},
  \bibinfo{pages}{014903} (\bibinfo{year}{1999}), \eprint{hep-ph/9902259}.

\bibitem[{\citenamefont{Lewin and Smith}(1996)}]{Lewin:1995rx}
\bibinfo{author}{\bibfnamefont{J.~D.} \bibnamefont{Lewin}} \bibnamefont{and}
  \bibinfo{author}{\bibfnamefont{P.~F.} \bibnamefont{Smith}},
  \bibinfo{journal}{Astropart. Phys.} \textbf{\bibinfo{volume}{6}},
  \bibinfo{pages}{87} (\bibinfo{year}{1996}).

\bibitem[{\citenamefont{Zyla et~al.}(2020)}]{Zyla:2020zbs}
\bibinfo{author}{\bibfnamefont{P.~A.} \bibnamefont{Zyla}} \bibnamefont{et~al.}
  (\bibinfo{collaboration}{Particle Data Group}), \bibinfo{journal}{PTEP}
  \textbf{\bibinfo{volume}{2020}}, \bibinfo{pages}{083C01}
  (\bibinfo{year}{2020}).

\bibitem[{\citenamefont{Angeli and Marinova}(2013)}]{Angeli:2013epw}
\bibinfo{author}{\bibfnamefont{I.}~\bibnamefont{Angeli}} \bibnamefont{and}
  \bibinfo{author}{\bibfnamefont{K.~P.} \bibnamefont{Marinova}},
  \bibinfo{journal}{Atom. Data Nucl. Data Tabl.} \textbf{\bibinfo{volume}{99}},
  \bibinfo{pages}{69} (\bibinfo{year}{2013}).

\bibitem[{\citenamefont{Piekarewicz et~al.}(2016)\citenamefont{Piekarewicz,
  Linero, Giuliani, and Chicken}}]{Piekarewicz:2016vbn}
\bibinfo{author}{\bibfnamefont{J.}~\bibnamefont{Piekarewicz}},
  \bibinfo{author}{\bibfnamefont{A.~R.} \bibnamefont{Linero}},
  \bibinfo{author}{\bibfnamefont{P.}~\bibnamefont{Giuliani}}, \bibnamefont{and}
  \bibinfo{author}{\bibfnamefont{E.}~\bibnamefont{Chicken}},
  \bibinfo{journal}{Phys. Rev. C} \textbf{\bibinfo{volume}{94}},
  \bibinfo{pages}{034316} (\bibinfo{year}{2016}), \eprint{1604.07799}.

\bibitem[{\citenamefont{Papoulias et~al.}(2020)\citenamefont{Papoulias, Kosmas,
  Sahu, Kota, and Hota}}]{Papoulias:2019lfi}
\bibinfo{author}{\bibfnamefont{D.~K.} \bibnamefont{Papoulias}},
  \bibinfo{author}{\bibfnamefont{T.~S.} \bibnamefont{Kosmas}},
  \bibinfo{author}{\bibfnamefont{R.}~\bibnamefont{Sahu}},
  \bibinfo{author}{\bibfnamefont{V.~K.~B.} \bibnamefont{Kota}},
  \bibnamefont{and} \bibinfo{author}{\bibfnamefont{M.}~\bibnamefont{Hota}},
  \bibinfo{journal}{Phys. Lett. B} \textbf{\bibinfo{volume}{800}},
  \bibinfo{pages}{135133} (\bibinfo{year}{2020}), \eprint{1903.03722}.

\bibitem[{\citenamefont{Stodolsky}(1966)}]{Stodolsky:1966zz}
\bibinfo{author}{\bibfnamefont{L.}~\bibnamefont{Stodolsky}},
  \bibinfo{journal}{Phys. Rev.} \textbf{\bibinfo{volume}{144}},
  \bibinfo{pages}{1145} (\bibinfo{year}{1966}).

\bibitem[{\citenamefont{Freedman}(1974)}]{Freedman:1973yd}
\bibinfo{author}{\bibfnamefont{D.~Z.} \bibnamefont{Freedman}},
  \bibinfo{journal}{Phys. Rev.} \textbf{\bibinfo{volume}{D9}},
  \bibinfo{pages}{1389} (\bibinfo{year}{1974}).

\bibitem[{\citenamefont{Kopeliovich and Frankfurt}(1974)}]{Kopeliovich:1974mv}
\bibinfo{author}{\bibfnamefont{V.~B.} \bibnamefont{Kopeliovich}}
  \bibnamefont{and} \bibinfo{author}{\bibfnamefont{L.~L.}
  \bibnamefont{Frankfurt}}, \bibinfo{journal}{JETP Lett.}
  \textbf{\bibinfo{volume}{19}}, \bibinfo{pages}{145} (\bibinfo{year}{1974}).

\bibitem[{\citenamefont{Freedman et~al.}(1977)\citenamefont{Freedman, Schramm,
  and Tubbs}}]{Freedman:1977xn}
\bibinfo{author}{\bibfnamefont{D.~Z.} \bibnamefont{Freedman}},
  \bibinfo{author}{\bibfnamefont{D.~N.} \bibnamefont{Schramm}},
  \bibnamefont{and} \bibinfo{author}{\bibfnamefont{D.~L.} \bibnamefont{Tubbs}},
  \bibinfo{journal}{Ann. Rev. Nucl. Part. Sci.} \textbf{\bibinfo{volume}{27}},
  \bibinfo{pages}{167} (\bibinfo{year}{1977}).

\bibitem[{\citenamefont{Hoferichter et~al.}(2020)\citenamefont{Hoferichter,
  Men\'endez, and Schwenk}}]{Hoferichter:2020osn}
\bibinfo{author}{\bibfnamefont{M.}~\bibnamefont{Hoferichter}},
  \bibinfo{author}{\bibfnamefont{J.}~\bibnamefont{Men\'endez}},
  \bibnamefont{and} \bibinfo{author}{\bibfnamefont{A.}~\bibnamefont{Schwenk}},
  \bibinfo{journal}{Phys. Rev. D} \textbf{\bibinfo{volume}{102}},
  \bibinfo{pages}{074018} (\bibinfo{year}{2020}), \eprint{2007.08529}.

\bibitem[{\citenamefont{Aristizabal~Sierra
  et~al.}(2023)\citenamefont{Aristizabal~Sierra, Barrow, Dutta, Kim, Strigari,
  Snowden-Ifft, and Wood}}]{AristizabalSierra:2022jgg}
\bibinfo{author}{\bibfnamefont{D.}~\bibnamefont{Aristizabal~Sierra}},
  \bibinfo{author}{\bibfnamefont{J.~L.} \bibnamefont{Barrow}},
  \bibinfo{author}{\bibfnamefont{B.}~\bibnamefont{Dutta}},
  \bibinfo{author}{\bibfnamefont{D.}~\bibnamefont{Kim}},
  \bibinfo{author}{\bibfnamefont{L.}~\bibnamefont{Strigari}},
  \bibinfo{author}{\bibfnamefont{D.}~\bibnamefont{Snowden-Ifft}},
  \bibnamefont{and} \bibinfo{author}{\bibfnamefont{M.~H.} \bibnamefont{Wood}}
  (\bibinfo{collaboration}{\ensuremath{\nu}BDX-DRIFT}), \bibinfo{journal}{Phys.
  Rev. D} \textbf{\bibinfo{volume}{107}}, \bibinfo{pages}{013003}
  (\bibinfo{year}{2023}), \eprint{2210.08612}.

\bibitem[{\citenamefont{Akimov et~al.}(2018)}]{Akimov:2018ghi}
\bibinfo{author}{\bibfnamefont{D.}~\bibnamefont{Akimov}} \bibnamefont{et~al.}
  (\bibinfo{collaboration}{COHERENT}) (\bibinfo{year}{2018}),
  \eprint{1803.09183}.

\end{thebibliography}
\end{document}